\documentclass[twocolumn, aps, nofootinbib]{revtex4-2}

\DeclareUnicodeCharacter{0301}{*************************************}

\usepackage{amssymb,amsmath,amsfonts,empheq,amsbsy,epsfig,wrapfig,float,slashed,graphicx, pstricks}
\usepackage{systeme}

\newcommand{\eq}[2]{\begin{equation}\label{#1}#2 \end{equation}}

\newcommand{\da}[1]{\left\langle \hspace{-2pt}\left\langle #1 \right\rangle \hspace{-2pt}\right\rangle}

\newcommand{\Tr}[1]{{\rm Tr}}

\usepackage{xcolor}

\def\bea{\begin{eqnarray}}
	\def\eea{\end{eqnarray}}
\def\be{\begin{equation}}
	\def\ee{\end{equation}}
\def\ba{\begin{array}}
	\def\ea{\end{array}}

\def\nn{\nonumber}

\begin{document}
	
\title{Renormalization of Interacting Random Graph Models}
\author{Alessio Catanzaro$^{1,2,3}$}
\email{catanzaro@lorentz.leidenuniv.nl}
\author{Diego Garlaschelli$^{1,2}$}
\email{garlaschelli@lorentz.leidenuniv.it}
\author{Subodh P. Patil$^1$}
\email{patil@lorentz.leidenuniv.nl}

\affiliation{1.) Instituut-Lorentz for Theoretical Physics, {Leiden University, 2333 CA Leiden, The Netherlands}\\}
\affiliation{2.) IMT School for Advanced Studies, Piazza San Francesco 19, 55100 Lucca, Italy\\}
\affiliation{3.) Dipartimento di Fisica e Chimica, Universit\'a di Palermo,  Viale delle Scienze, Ed. 17, 90133 Palermo, Italy\\}
\date{\today}

\begin{abstract}
Random graphs offer a useful mathematical representation of a variety of real world complex networks. Exponential random graphs, for example, are particularly suited towards generating random graphs constrained to have specified statistical moments. In this investigation, we elaborate on a generalization of the former where link probabilities are conditioned on the appearance of other links, corresponding to the introduction of interactions in an effective generalized statistical mechanical formalism. When restricted to the simplest non-trivial case of pairwise interactions, one can derive a closed form renormalization group transformation for maximum coordination number two on the corresponding line graph. Higher coordination numbers do not admit exact closed form renormalization group transformations, a feature that paraphrases the usual absence of exact transformations in two or more dimensional lattice systems. We introduce disorder and study the induced renormalization group flow on its probability assignments, highlighting its formal equivalence to time reversed anisotropic drift-diffusion on the statistical manifold associated with the effective Hamiltonian. We discuss the implications of our findings, stressing the long wavelength irrelevance of certain classes of pair-wise conditioning on random graphs, and conclude with possible applications. These include modeling the scaling behavior of preferential effects on social networks, opinion dynamics, and reinforcement effects on neural networks, as well as how our findings offer a systematic framework to deal with data limitations in inference and reconstruction problems. 
\end{abstract}
\maketitle

\section{Preliminary remarks}
One of the most utilized approaches in modeling and studying the properties of random networks is the so-called Exponential Random Graph (ERG) framework \cite{doi:10.1137/1028156, HJ_1999, PhysRevE.70.066117, PN_two_star,  Park_2005, coolen2017generating, squartini2017maximum, cimini2019statistical}. Its formulation builds on the realization that in a precise sense, the formalism of statistical physics is ontologically equivalent to statistical inference, and can be formalized more broadly in terms of a \textit{generalized statistical mechanics} \cite{PhysRev.106.620, jaynes1957information}. The workhorse of this generalization lies in viewing the Gibbs entropy formula as a use case of Shannon information entropy \cite{shannon}, which facilitates applications beyond physical systems towards any system where observable quantities are drawn from a random ensemble. The implications of this are as immediate as they are profound. One can define a probability measure that reproduces the observed statistics of any given random system and associate it with the `Hamiltonian' of a generalized statistical mechanical ensemble. This is done by demanding that the measure replicate all observed moments in a manner that maximizes its Shannon information, and therefore being minimally presumptive of any other priors\footnote{In his seminal work, E.T. Jaynes credits Shannon as supplying the missing ingredient required to actualize Laplace's \textit{Principle of Insufficient Reason} for assigning probabilities to account for all possible draws from a random ensemble constrained to reproduce a fixed set of moments, and possibly supplemented with additional structural priors \cite{PhysRev.106.620, jaynes1957information}.}. One recovers the usual Boltzmann weight when applied to thermodynamic systems, and in the context of random graphs, the result is the formalism of exponential random graphs \cite{doi:10.1137/1028156, PhysRevE.70.066117}.\\

Given that the properties of a random graph can be derived from its adjacency matrix, the ERG formalism associates a Hamiltonian to any random graph ensemble such that the probability of realizing any given draw for the adjacency matrix $A_{ij}$ is given by 
\begin{equation}
	P(\textbf{A}) \equiv  \frac{e^{-H(\textbf{A})}}{Z};~~~~ Z := \sum_{\{A_{ij} \}} e^{-H(\textbf{A})}, \label{eq:erg}
\end{equation}
where $Z$ is referred to as the partition function, and where the Hamiltonian is in general a multilinear function of the elements of the adjacency matrix:
\begin{multline}
	H(\textbf{A}) =  - h^{(0)} - \sum_{ij} h^{(1)}_{ij} A_{ij} - \sum_{ij\neq kl} h^{(2)}_{ijkl} A_{ij}A_{kl}\\
	- \hspace{-10pt}\sum_{\substack{ij \neq kl,\\ mn 
			kl \neq mn}} h^{(3)}_{ijklmn} A_{ij}A_{kl}A_{mn} - ... \label{eq:ham}
\end{multline}
The form of the Hamiltonian parameterizes generic interactions among links related to various graph observables, such as the number and distribution of triangles or wedges. Its couplings up to some order can be fixed by demanding expectation values associated with the probability assignment (defined by Eq. \ref{eq:erg}) for a sufficient number of observables $\{\mathcal O_a\}$ be constrained to assigned values:
\eq{}{\langle \mathcal O_a\rangle = \sum_{\textbf{A}}P(\textbf{A})\mathcal O_a(\textbf{A}),}
subject to maximization of the Gibbs-Shannon entropy
\eq{}{S = -\sum_{\textbf{A}}P(\textbf{A})\log P(\textbf{A}).}
In practice, one only ever has access to a finite number of observables, themselves subject to sample variance and potential estimator bias. This will not be an issue for the intents and purposes of a predictive framework for reasons that we will elaborate upon shortly.\\   

The $\{A_{ij}\}$ sum over the set $\{0,1\}$ for unweighted graphs, and otherwise integrate over arbitrary values for weighted graphs. In what follows, we restrict ourselves to the unweighted case. The simplest example of the latter is furnished by the Erd\H{o}s-Renyi random graph, for which all couplings in Eq. \ref{eq:ham} vanish except for $h^{(1)}_{ij} = h^{(1)}_{ji} = \log[p/(1-p)]$ for $i \neq j$ (see e.g. \cite{catanzaro2025}). A disordered variant of the Erd\H{o}s-Renyi graph is obtained by allowing the $h^{(1)}_{ij}$ to vary link-wise, so that
\eq{hdef}{h^{(1)}_{ij} = \log\left(\frac{p_{ij}}{1-p_{ij}}\right).}

The exponential random graph approach has been successful in a number of contexts, including network inference problems, and is considered as the benchmark for null models of random graphs. However, this approach comes up against a number of difficulties in wider application. One is that while straightforwardly solvable for the separable case --- that is, when 
\eq{eq:fac}{Z \equiv Z_1 Z_2 ... Z_{\mathcal N},~~~~ \mathcal N := N(N-1)/2,}
where each individual factor corresponds to the partition function for each possible individual link\footnote{For example, the partition function for a graph with $N$ nodes factorizes as per Eq. \ref{eq:fac} when one truncates the expansion beyond terms linear in $A_{ij}$ in Eq. \ref{eq:ham}.} --- when correlations and higher order conditionings are introduced, analytic solutions are no longer so straightforwardly obtainable. Moreover, in phenomenological approaches where one attempts to reconstruct graph Hamiltonians that reproduce various features inferred from data, one is not only hindered by sample variance, but also by data that is limited to network realizations of specific sizes. Certain features corresponding to graph generative processes, i.e. involving the dynamics of graph formation and growth, might elude an inference approach based on the standard formulation of the exponential random graph framework.\\

Perhaps the best known mechanism of graph formation in network science is furnished by the Barabasi-Albert network model, in which a `rich-get-richer' process drives graph evolution towards particular asymptotic scaling properties \cite{Baraba_si_1999, Albert_2002}. The conceptual utility of these models derives from their simplicity, which also inevitably limits their applicability in modeling complex systems where multiple processes can shape connections. To  account for the latter, one might consider introducing dynamical features that depend on various building blocks of the graph. One such class of models are often referred to as fitness (or hidden variable) models \cite{caldarelli2002scale, garlaschelli2004fitness, van2016random, van2024random, cimini2019statistical, catanzaro2025spectra} where one assigns node-specific variables, with graph interactions dependent on these quantities. The assignment and distribution of the fitness variables end up determining the macroscopic properties of the graph and its associated network structures, demanding answers to questions such as which features are relevant at the largest scales, which are irrelevant, and how the presence of disorder in their assignment is reflected in the statistics of macroscopic observables. 

\subsection{The (effective) generalized statistical mechanics of random graphs}

In what follows, we develop the means to address some of the aforementioned issues by taking the generalized statistical mechanical framework in which the ERG formalism originates to its natural completion, from both a formal and an inferential point of view. We do this by embedding it within the epistemic framework furnished by the paradigm of \textit{effective theories} \cite{Burgess_2020, Manohar:2018aog, shankar1997, Schakel:2008zz}. The idea is simple to state -- one simply views the Hamiltonian in Eq. \ref{eq:ham} as a moment generating functional that has been operator Taylor expanded without loss of generality. Each specification of the coefficients in its expansion parameterizes a specific point in theory space, where the only prior restrictions on the coefficients stem from the symmetries and structural constraints that define the system.\\ 

Although at first glance it might seem that one has to specify a tower of coefficients in order to have a predictive or retrodictive framework, this is operationally not the case given that one only ever has access to limited resolution and sample sizes in observations. Consequently, only a finite number of coefficients have to be fixed (which is to work up to a certain order in the operator expansion) in order to have a predictive framework for all subsequent observations to the stipulated accuracy\footnote{\label{fn:wc}This presumes that one works in an operator basis such that the analog of Eq. \ref{eq:ham} represents an expansion whose leading order terms correspond to an exactly solvable system (however non-linear or strongly correlated), where the higher order terms parameterize collective modes that perturb away from this solution in a convergent series.}. The reason for this can be understood both from a renormalization group (RG) and a statistical inference point of view. When viewed as an \textit{effective} Hamiltonian, the terms in the expansion of Eq. \ref{eq:ham} represent an ordered series in terms of relevance: \textit{Power counting relevance} from the perspective of a renormalization group analysis, and \textit{parameter relevance} from the perspective of statistical inference. The former can be viewed as a special case of the latter through the portal of Fisher information \cite{Machta_2013,quinn2022}, in a manner that is straightforward to articulate.\\

Sequentially higher order terms in the effective Hamiltonian (Eq. \ref{eq:ham}) have an increasingly smaller effect on observable quantities, caveating footnote \ref{fn:wc}. That is, attempting to fix the parameters of the Hamiltonian with inevitably limited data will result in a likelihood function that will exhibit progressively weaker parametric dependence on higher order couplings in a pronounced hierarchy. Leading order couplings will correspond to the largest eigenvalues of the associated Fisher-Rao information metric \cite{rao1945information, e22101100, amari2000methods} (reviewed in section \ref{sec:CD}), or have the greatest parameter relevance. Higher order couplings will correspond to successively smaller eigenvalues, or be increasingly \textit{ir}relevant. This is perhaps best illustrated through an example that ought to be familiar to many readers.\\

For continuum physical systems, the renormalization group notion of power counting relevance translates directly into a derivative expansion. By expanding in a series of derivative operators that is consistent with the symmetries of the system (e.g. rotational symmetry), one can effectively model arbitrary dynamical processes up to some resolution. Since derivatives have physical dimension, successively higher derivative terms require compensating dimensional factors in order to book keep the expansion. For example, one can consider modeling an isotropic diffusion process via the following effective equations: 
\eq{eq:diff}{\frac{d \rho}{dt} = \sigma\rho + D_2\nabla^2\rho + D_4 \ell^2  \nabla^4 \rho + D_6 \ell^4 \nabla^6 \rho + ...} 
where as written, all the higher order coefficients $D_4, D_6$ etc. have the same physical dimension as the diffusion coefficient $D_2$. At spatial resolutions $L \gg \ell $, so that the Fourier modes associated with these scales have wave numbers that satisfy $k_L/k_\ell \ll 1$, we see that Eq. \ref{eq:diff} represents a convergent operator Taylor expansion where successive terms are subleading in even powers of $k_L/k_\ell$. We'll return to the physical significance of the so far arbitrary scale $\ell$ shortly.\\

In order for the quartic operator to have a comparable effect to the leading diffusion term, the $D_4$ coefficient would have to be dialed up to $D_4 \sim D_2 k_\ell^2/k_L^2 \gg D_2$. Similarly, one would need $D_6 \sim D_4 k_\ell^2/k_L^2 \gg D_4$ in order for the sextic operator to compete with the quartic, and so on. Conversely, unless the $D_4$ coefficient starts to approach $D_4 \approx D_2 k_\ell^2/k_L^2$, the quartic operator will have negligible effects on long wavelength observables in comparison to the quadratic diffusion operator. In fact, there is a wide range within the parameter interval defined by the condition $|D_4| \ll |D_2 k_\ell^2/k_L^2|$ where $D_4$ can vary and have parametrically negligible effects on long wavelength observables relative to the leading diffusion term. That is, the likelihood function constructed from repeated observation will have a much weaker dependence on $D_4$ than $D_2$, and will hence be associated with a smaller eigenvalue of the associated Fisher information metric.\\

In the context of a renormalization group analysis, when one successively coarse grains over small scales and re-expands the effective equations of motion as in Eq. \ref{eq:diff} after each coarse graining, one will find that the value of $D_2$ remains roughly the same, whereas $D_4$ decreases. For this reason, $D_2$ is classified as \textit{marginally relevant}, and $D_4$ is classified as \textit{irrelevant}, with both notions exactly corresponding to their parameter relevance as determined by their associated Fisher metric eigenvalues \cite{Machta_2013, quinn2022, transtrum2015perspective}\footnote{The classification of operators into categories of relevance, marginal relevance, and irrelevance under the action of the renormalization group is owed to Kadanoff and Wegner \cite{Kadanoff1976, wegner1976critical}. }. All that remains is to clarify the nature of the scale $\ell$ in Eq. \ref{eq:diff}: In a precise operational sense, it is the resolution scale of our effective description. Were one to average over all fluctuations with wavelengths smaller than the length scale $\ell$, the resulting equations of motion for \textit{all} modes with wavelengths longer than than $\ell$ would be reproduced by an operator expansion corresponding to Eq. \ref{eq:diff}, parametrically weighted by factors of $\ell^2$ with order unity numerical prefactors. If one views diffusion processes as having coarse grained an underlying kinetic theory description \cite{landau1981physical,succi2013lattice}, one can explicitly demonstrate this via an influence kernel parameterization of inter-particle forces (see e.g. \cite{Grosvenor:2024vcn} in the context of broken symmetries).\\

The realization that one can parameterize one's ignorance of small scale processes via an operator expansion and still end up with a tractable and predictive framework for macroscopic observables is what freed particle physicists from having to work within the confines of \textit{renormalizable} field theories\footnote{\label{fn}In particle physics effective field theory, the adjective \textit{renormalizable} is reserved for those theories which form a closed subspace under renormalization group flow in the space of all theories, parameterized via some operator basis. Effective theories in general need not be renormalizable by this definition, but nevertheless remain predictive as discussed, as only a finite subset of operators corresponding to the leading order terms in any derivative expansion remain relevant or marginal as one flows to the infra-red (see e.g. \cite{Burgess_2020, Manohar:2018aog}).}, and ushered in the era of \textit{effective field theories}. The latter represents the state of the art in particle phenomenology, and it features an underlying logic that transcends its original application towards continuum systems. One simply writes down all possible operators that could appear in the moment generation functional not forbidden by symmetries or other kinematic or structural constraints (i.e. the effective action in the context of particle physics, or the effective Hamiltonian in the context of a generalized statistical mechanical system), and classifies them according to their parameter relevance.

\subsection{Universality classes of graph Hamiltonians}

It is from the vantage of effective theories that renormalization group analyses take on a categoric importance of their own. Not only do RG methods give us the tools to determine parameter relevance, their complete implementation allows us to unambiguously determine which quantities are legitimate observables at long wavelengths, and what may be artifacts of various approximation schemes. An important corollary of this classification is that all models flow under renormalization towards a handful of equivalence classes at large scales, and given the special designation of \textit{universality classes}. Which universality class one flows to is determined entirely by the invariances or symmetries of the system, along with how these symmetries are broken as one changes scale.\\

In what follows, we attempt the first steps in understanding the behavior of a generalized statistical mechanical parameterization of unweighted random graph models under renormalization group flow. Our starting point is the operator expansion Eq. \ref{eq:ham}, whose most immediate consequence is that the probability for any link to appear will generically be conditioned on the appearance of other links. As a consequence, the partition function for the graph will not be factorizable into a product of independent partition functions for each link. That is, 
\eq{}{ Z \neq Z_1 Z_2 ... Z_{\mathcal N},}
where once again $\mathcal N$ counts the number of links in a graph with $N$ nodes, and we omit the possibility of loops (edges that begin and end on the same node) in the discussion to follow. The simplest non-trivial case of this is when one restricts to bilinear interactions, which can be written without loss of generality as:
\eq{Ham}{   H(\textbf{A}) = -\sum_{i\leq j} \varepsilon_{ij} A_{ij} - \frac{1}{2} \sum_{\substack{i \leq j, k \leq l\\ (i,j)\neq(k,l)}} \beta_{ijkl} A_{ij} A_{kl},}
where the pairs $(i,j)$ and $(k,l)$ denote the links connecting the respectively indexed nodes. Tripartite and higher order conditioning can similarly be introduced via cubic and higher order interactions via a generating function formalism, which we review in Appendix \ref{app:A}. If this higher order conditioning is weak enough, one can avail of the diagrammatic methods of statistical field theory and work in a perturbative series in the corresponding small cubic and higher couplings (see e.g. \cite{Helias_2020}). Although we won't pursue this possibility in the present investigation, a more complete study of parameter relevance in random graph models would necessitate it. Instead, we focus only on bilinear interactions in what follows and defer a more comprehensive study to a future investigation. A rich structure already emerges within this class, where moreover, the existence of a closed subspace under exact RG transformations can be demonstrated.\\

Truncating the effective Hamiltonian to bilinear interactions might seem like an academic oversimplification at first glance. However, as elaborated in the next section, Eq. \ref{eq:ham} can also be viewed as the Hamiltonian of an equivalent disordered spin system with couplings of arbitrary order (i.e. in terms of degrees of nearest neighbors), whose effective dimensionality is determined by the number of nearest neighbors each site has. We refer to this quantity as the coordination number of a link, which is encoded in all the $h^{(2)}_{ijkl}$ terms being different from zero. Were one to presume a hierarchy for these terms over higher order couplings,  the result of successive RG transformations for a large class of bare effective Hamiltonians would be to flow towards the form Eq. \ref{Ham}, which therefore represents a universality class of equivalent random graph models, and therefore an important object of study on its own terms. Moreover, from a practical and application intensive point of view, Eq. \ref{Ham} can also effectively model temporal Markov network processes, where RG transformations would correspond to temporal coarse graining. We elaborate on this in more detail in an accompanying study \cite{Nextpaper}, and commence our formal study of random graph renormalization in the following sections. 

\subsection{Outline}

The goal of this paper is to set up a general framework through which one might quantify the scaling of interactions in random graph models. That is, to determine the relevance, or irrelevance of the conditionings of different graph variables on each other. Although this is highly non-trivial problem in full generality, we focus in what follows on a restricted class of transformations for which closed form renormalization group transformations are available, which we argue to represent a universality class of graph Hamiltonians.\\

In the sections that follow, we first set up the formalism of random graph renormalization in the line graph representation in section \ref{sec:LG}, elaborating on the existence of closed form RG transformations for maximum coordination number two pairwise interactions. We then consider how these interactions renormalize when the couplings are homogeneously assigned in \ref{sec:HRG}, and generalize to the disordered setting in \ref{sec:RD}, where we immediately find that the problem can be recast in terms of determining the the induced flow on the disorder probability assignments. We understand this flow to be akin to time reversed drift diffusion, as detailed in appendix \ref{sec:CD}, with additional details relevant to our discussion collected in appendices \ref{app:A} \ref{app:NM}. We conclude our investigation with a discussion of the implications of our findings, the most pertinent of which is that within the universality class of bilinear graph Hamiltonians, all models are \textit{infra-red} trivial, meaning that all pairwise conditions flow to irrelevance at large enough scales. We translate the ramifications of this for a number of applications, and close with a discussion of open questions and future directions. 

\section{\label{sec:LG} Line graphs, spin chains, and closed RG transformations}
The graph Hamiltonian Eq. \ref{Ham} can be re-expressed as the Hamiltonian of a disordered spin chain via mapping to the line graph: 
\eq{LineHam}{H(\textbf{A}) = -\sum_{a} \varepsilon_{a} A_{a} - \frac{1}{2} \sum_{a\leq b} \beta_{ab} A_{a} A_{b},}
where distinct pairs of indices $(i, j)$ are mapped to a single index $a$, so that edges have been mapped to nodes, and where $1 \leq a  \leq \mathcal N$. We stress that a priori, there is no natural notion of nearest neighbor on a random graph at the level of the effective Hamiltonian Eq. \ref{LineHam}. Metric notions of nearest neighbors and local neighborhoods -- key to the usual implementation of the renormalization group -- are only something that can be defined for a specific instance of a random graph\footnote{For a recent review on progress and open issues regarding network renormalization, see \cite{gabrielli2025network}.}. Moreover, the couplings $\varepsilon_a$ and $\beta_{ab}$ are a priori completely arbitrary, and need not even correspond to `local' interactions. That is, without additional restrictions, the probability of any given link appearing can in principle be conditioned on an arbitrary number of other links.\\

Nevertheless, a class of bilinear interactions can be distinguished by the fact that they remain closed under an exact RG transformation that corresponds to decimation on the line graph. Specifically, we consider interactions of the form:
\eq{cn2}{\beta_{ab} = g_a \delta_{a,b-1} + g_{b}\delta_{a,b+1},}
so that
\eq{GH}{ H(\textbf{A}) = - \sum_{a=1}^{\mathcal N} \left[\varepsilon_{a} + g_a A_{a+1}\right]A_a,}
which formally looks like the Hamiltonian of a paramagnetic spin glass with nearest neighbor interactions, in spite of the absence of such notions on a random graph. The coefficients $\varepsilon_a$ and $g_a$ themselves could be drawn from independent ensembles, corresponding to arbitrary `local' disorder.\\ 

As discussed in Appendix \ref{app:A}, the formal process of integrating out one or more of the edge degrees of freedom in the context of Eq. \ref{GH} is equivalent to statistical marginalization, and effects a decimation renormalization group transformation on the line graph. It should be stressed at the outset that there is no a priori compulsion to have imposed this transformation uniformly, and one can consider determining the parameters of the effective Hamiltonian obtained by marginalizing over a single link in isolation (in \cite{Nextpaper} we take an approach to network renormalization that is premised on iterating this procedure). The result of having decimated the $n$'th link is a locally coarse-grained line graph with an effective Hamiltonian with the following renormalized couplings\footnote{Up to an overall rescaling of the partition function that is equivalent to a localized additive free energy contribution given by $F_n = \log\left[1 + \varepsilon_n\right]$, corresponding to the individual contribution of the marginalized over link in Eq. \ref{FE}.}:
\begin{eqnarray}
	\label{RG1g}
	\widetilde g_{n-1} &=& \log\left\{\frac{\left(1 + e^{g_{n-1} + g_n +  \varepsilon_{n}}\right)\left(1 + e^{ \varepsilon_{n}}\right)}{\left(1 + e^{g_n + \varepsilon_{n}}\right)\left(1 + e^{g_{n-1}+ \varepsilon_{n}}\right)}\right\},\\
	\nonumber
	\widetilde \varepsilon_{n-1} &=& \varepsilon_{n-1} + \log\left\{\frac{\left(1 + e^{g_{n-1} + \varepsilon_{n}}\right)}{\left(1 + e^{\varepsilon_{n}}\right)}\right\},
	\\
	\nonumber
	\widetilde \varepsilon_{n+1} &=& \varepsilon_{n+1} + \log\left\{\frac{\left(1 + e^{g_n + \varepsilon_{n}}\right)}{\left(1 + e^{\varepsilon_{n}}\right)}\right\},
\end{eqnarray}
with all others left unchanged. On the other hand, the effect of uniform decimation, where every alternate node on the line graph is successively marginalized over is to effect the following transformations:
\begin{eqnarray}
	\label{RGg}
	\widetilde g_n &=& \log\left\{\frac{\left(1 + e^{g_n + g_{n+1} + \varepsilon_{n+1}}\right)\left(1 + e^{ \varepsilon_{n+1}}\right)}{\left(1 + e^{g_n + \varepsilon_{n+1}}\right)\left(1 + e^{g_{n+1}+ \varepsilon_{n+1}}\right)}\right\},\\ \nonumber
	\widetilde \varepsilon_n &=& \varepsilon_n + \log\left\{\frac{\left(1 + e^{g_n + \varepsilon_{n+1}}\right)\left(1 + e^{g_{n-1} + \varepsilon_{n-1}}\right)}{\left(1 + e^{\varepsilon_{n+1}}\right)\left(1 + e^{\varepsilon_{n-1}}\right)}\right\}
\end{eqnarray}
where we've relabeled the indices to run over the integers again after each successive uniform decimation. A simple derivation of these equations can be found in the context of the Ising Model with sparse disorder (see e.g. \cite{Bar-Yam:2018ytq}), or in our accompanying paper \cite{Nextpaper}.\\

The closure of these RG transformations -- meaning that each RG step returns an effective Hamiltonian of the same operator form up to rescaled couplings and an overall rescaling of the partition function -- results from the fact that Eq. \ref{cn2} represents bilinear interactions of maximum coordination number two in the line graph representation. Consequently, the associated Hamiltonian Eq. \ref{GH} maps onto that of a one dimensional spin glass with localized disorder and an inhomogeneous external magnetic field, inheriting the closure and exact implementability of renormalization group transformations specific to one dimensional systems. This is no longer the case the instance we add even a single interaction with coordination number three or higher, with successive RG transformations driving the system towards highly heterogenous all-to-all couplings (the homogeneous limit of which would be a Curie-Weiss like model).\\
\begin{figure}[t]
	\centering
	\includegraphics[width=\linewidth]{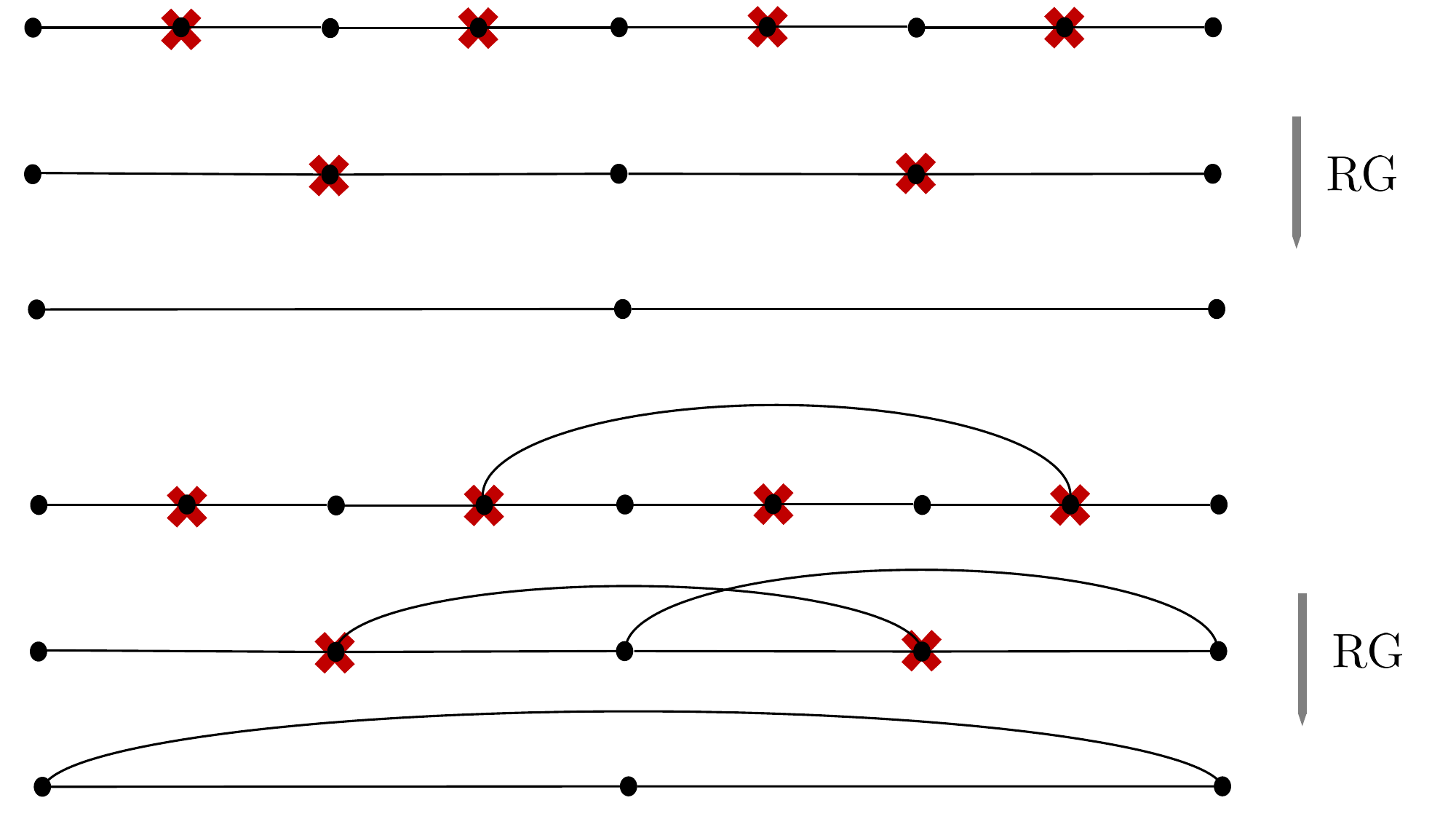}
	\caption{Schematic rendering of maximum coordination two couplings (top) and a single bare coordination number three coupling (bottom) under RG transformations. The red crosses indicate the sites that are decimated at each RG step. More generally, successive RG steps increase the maximum coordination number, and flow towards all-to-all couplings (bottom most diagram).}
	\label{fig:RG}
\end{figure}

Specifically, consider replacing some of the terms in Eq. \ref{GH} with the following coordination number three interaction:  
\eq{}{H \supset \left[\varepsilon_{m} + g_m A_{m+1} + g'_m A_{m+k}\right]A_a}
for some fixed $m$ and some integer $k$. In the context of the analog lattice system where one can avail of notions of metric distance, $k = 2$ would correspond to a next to next to nearest neighbor interaction. In the present context, we leave $k$ arbitrary, and view it as corresponding to conditioning the $m$'th and $m+k$'th links on each other. Under repeated decimation, the maximum coordination number on the line graph would successively increase in such a manner such that when one considers randomly distributing higher coordination interactions across the graph, the effective Hamiltonian would flow towards all-to-all couplings, schematically depicted in Fig. \ref{fig:RG}. 

\subsection{Maximum coordination number two $\equiv$ disordered spin chain}

The fact that maximum coordination number two and all-to-all couplings represent the two universality classes that admit closed form renormalization group transformations is a transcription of the closure of RG transformations in one dimensional lattice spin systems to the context of random graph models. We can also make a straightforward identification for higher coordination numbers by noting that the lack of closure for RG transformations in general higher dimensional lattice systems necessitate approximation methods to force closure, such as bond shifting \cite{migdal1975phase, KADANOFF1976359}, variational, or mean field methods (see e.g. \cite{kadanoff2000statistical} for an overview)\footnote{Caveating the availability of exact solutions in the specific case of $d = 2$ in the absence of disorder \cite{onsager1944crystal, PhysRevB.19.2749}.}. In the absence of these approximations, summing over the contributions of short distance degrees of freedom generates higher order and beyond nearest neighbor interactions with each successive RG transformation. In spite of this, the availability of a predictive effective description via an effective Hamiltonian stems from the fact that these higher order and higher point interactions correspond to increasingly \textit{irrelevant} operators as one flows to the infra-red, and is the basis of the derivative expansion in continuum descriptions (cf. footnote \ref{fn}). Hence, RG transformations generate trajectories in theory space where all possible interactions are generated, albeit arranged in a hierarchy of relevance for long wavelength observables. The limit of this flow for an initial line graph with bilinear interactions of arbitrary coordination number is heterogeneous all-to-all couplings, unless the bare couplings form a closed subspace to begin with, as is the case for maximum coordination number two.

\section{\label{sec:HRG}Homogeneous RG flow}
\begin{figure}[t]
	\centering
	\includegraphics[width=0.85\linewidth]{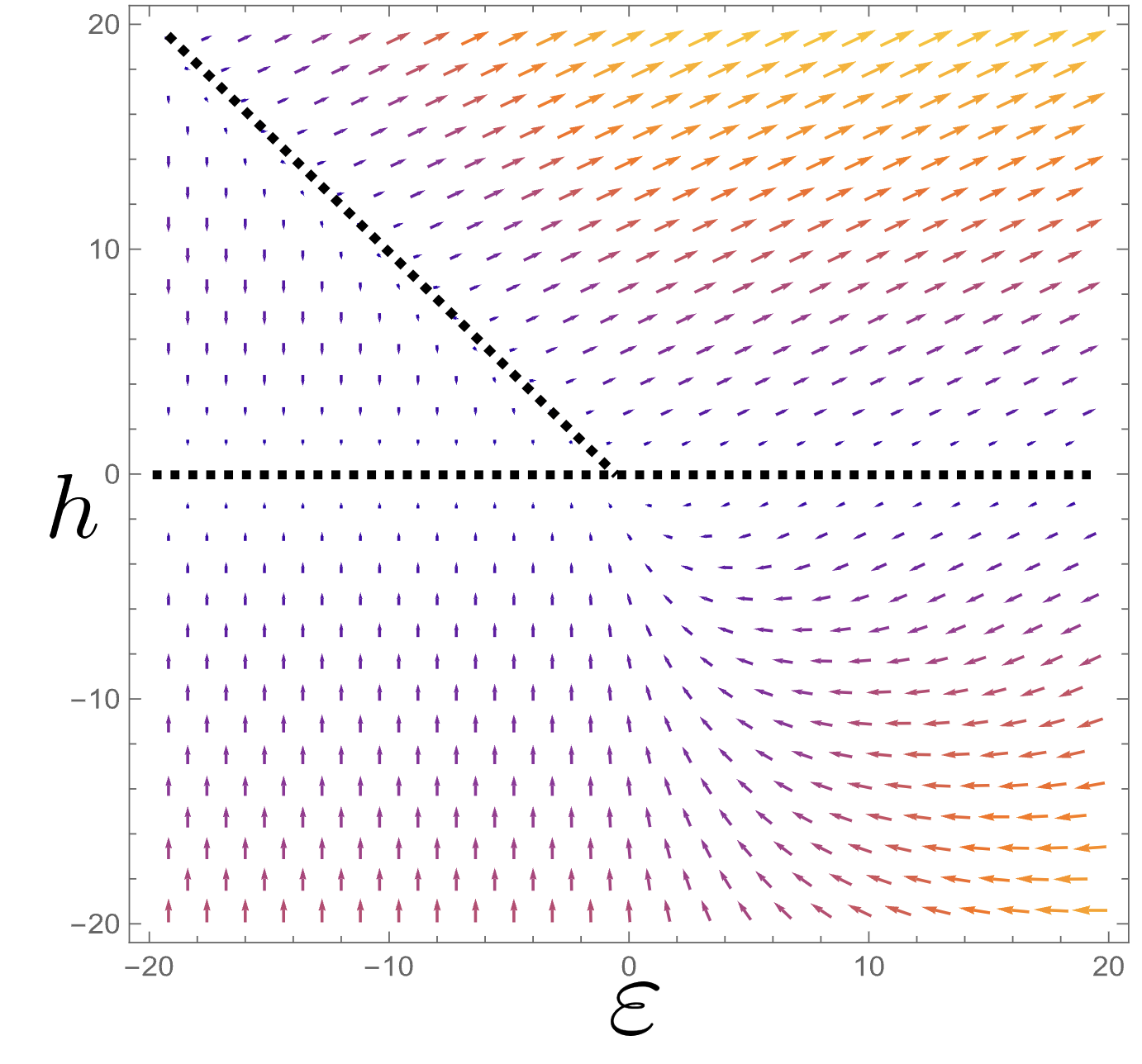}\\
	\vspace{10pt}
	\includegraphics[width=0.85\linewidth]{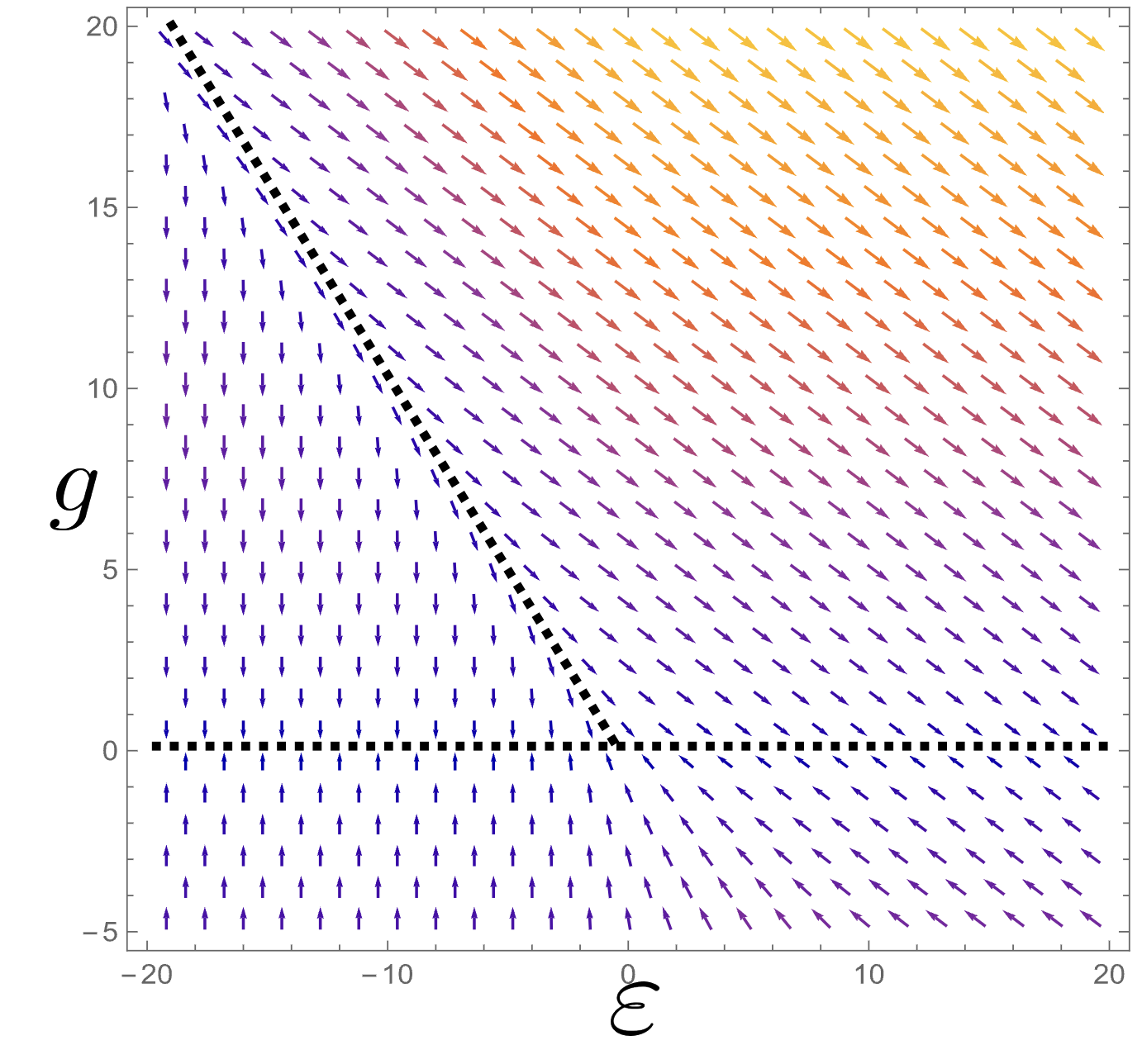}
	\caption{RG flow for homogeneous coordination number two interactions towards an Erd\H{o}s-Renyi fixed line, where arrows indicate direction of flow and color determine the its intensity, lighter being stronger. Upper panel: A shearing behavior for positively correlated conditioning is apparent, along with a separatrix along the line $\varepsilon = -g$ for $g>0$. This corresponds to the vanishing of the beta function of external magnetic field $h := \varepsilon + g$ along the line $h \equiv 0$ for the analog spin glass system, illustrated by the bottom panel.  }
	\label{fig:RG2}
\end{figure}

A simple but informative case to get our bearings with is the setting where each of the $\{g_a, \varepsilon_a\}$ couplings in Eq. \ref{GH} are respectively fixed to be the same, and we successively decimate, or marginalize over alternate links on the line graph (see for a thorough derivation \cite{Nextpaper}). In this case, Eqs. \ref{RGg} reduce to the following system  of equations: 
\begin{eqnarray}
\label{betaG}
\widetilde g &=& \log\left\{\frac{\left(1 + e^{2 g + \varepsilon}\right)\left(1 + e^{ \varepsilon}\right)}{\left(1 + e^{g + \varepsilon}\right)^2}\right\}, \\ \nonumber
\widetilde \varepsilon &=& \varepsilon + 2\log\left\{\frac{\left(1 + e^{g + \varepsilon}\right)}{\left(1 + e^{\varepsilon}\right)}\right\},
\end{eqnarray}
where tildes denote renormalized quantities, and where the discrete generalization of the beta functions, defined as $D g := \widetilde g - g \equiv\beta_g(g,\varepsilon)$, and $D \varepsilon := \widetilde \varepsilon - \varepsilon \equiv \beta_\varepsilon(g,\varepsilon)$, are given by:
\begin{eqnarray}
 \beta_g(g,\varepsilon) &=& \log\left\{\frac{\left(1 + e^{2 g + \varepsilon}\right)\left(1 + e^{ \varepsilon}\right)}{e^g\left(1 + e^{g + \varepsilon}\right)^2}\right\},\\ \beta_\varepsilon(g,\varepsilon) &=& 2\log\left\{\frac{\left(1 + e^{g + \varepsilon}\right)}{\left(1 + e^{\varepsilon}\right)}\right\}\nonumber.
\end{eqnarray}
The RG flow corresponding to these beta functions is depicted in Fig. \ref{fig:RG2}, where we see that successive transformations drive one towards the $g \equiv 0$ fixed line, so the links are no longer conditioned on each other and we recover an Erd\H{o}s-Renyi random graph at long wavelengths.

The various features of the flow diagram lend themselves to straightforward interpretation. We first note that the part of the diagram below the $g = 0$ axis is only meaningful for a single RG step. The reason is that implicit in Eq. \ref{betaG} is the fact that were we to begin with a negative value for the bare $g$, a single decimation iteration will return a positive value for $g$. Although the structure is far richer when we turn our attention to inhomogeneous and disordered interactions, the latter feature of considering only homogeneous interactions can be readily understood from the fact that on the analog spin system, negative values of $g$ corresponds to uniform anti-ferromagnetic couplings. When one considers the effect of marginalizing over two adjacent anti-ferromagnetic couplings, the result is a effective ferromagnetic coupling (i.e. $\widetilde g>0$) that maintains its ferromagnetic nature under further RG iterations.\\

The separatrix that is apparent along the line $\varepsilon = -g$ in the upper plane, along with the fixed line at $g=0$ can also be readily understood by referring to the analog spin system. Recalling Eq. \ref{GH} and making the transformation to the spin variables
\eq{}{A_a = \frac{1 + 2\sigma_a}{2},~~~~ \sigma_a = \pm \frac{1}{2},}
the result will be a Hamiltonian of the form:
\eq{SH}{H(\sigma) = - \sum_{a=1}^{\mathcal N} g_a \sigma_{a}\sigma_{a+1} - \sum_{a=1}^{\mathcal N} \sigma_{a}\left(\varepsilon_a + \frac{g_a + g_{a+1}}{2}\right)}
where we've neglected to include additive contributions to the free energy. When restricted to the case $\varepsilon_a \equiv \varepsilon$ and $g_a \equiv g$, we see that this Hamiltonian maps on to that of a one dimensional Ising model in the presence of an external magnetic field $h$ with the identification $h := \varepsilon + g$. Given that decimation in zero external field does not generate any one point terms in the renormalized Hamiltonian, the locus represented by $h \equiv 0$ or $\varepsilon = -g$ for represents an (unstable) fixed line in the upper $g$ plane, and an attractor in the lower $g$ plane\footnote{This is made explicit by considering the beta function for the linear combination $h = \varepsilon + g$, which is given by $\beta_h = \log\left\{\left(e^{g + \varepsilon} + e^{-g} \right)/\left(1 + e^{\varepsilon}\right)\right\}$, from which we see that $\beta_h \equiv 0$ on the locus $\varepsilon = -g$.}. Furthermore, the fact that the overall flow is driven to a fixed line at $g=0$ follows from fact that in one dimension, RG flow drives one towards the infinite temperature fixed point under the identification $g \equiv J/(k_BT)$, where $J$ denotes nearest neighbor couplings on the analog spin chain.\\  

Finally, the shearing structure on either side of this separatrix also admits a straightforward interpretation. When one positively or negatively conditions the probability of a given link to appear on the appearance of another link, it effectively increases, or respectively decreases the probability for that link to appear for homogeneous probability assignments, provided that the conditioning is strong enough. Therefore coarse graining will steadily drive the effective probability up or down depending on whether one is in the upper or lower $g$-plane until one is driven to the (analog infinite temperature) fixed line at $g \equiv 0$.\\

A far richer structure appears when one allows for the couplings to vary by site on the line graph. Although it would be meaningful to consider the effects of an arbitrary initial specification of the full set of couplings $\{\varepsilon_a,g_a\}$, a more general framing with an eye towards further applications would be to draw these from some probability distribution function (PDF), which is to introduce disorder on the line graph. 

\section{\label{sec:RD}Random disorder and induced probability flows}
A useful reformulation of the RG transformations Eqs. \ref{RGg} is in terms of the variables $y_n := e^{g_n}$ and $x_n := e^{\varepsilon_n}$, in which they take an algebraic form:
\begin{eqnarray}
	\label{Ybn}
	\widetilde y_n &=& \frac{(1 + y_n y_{n+1}x_{n+1})(1 + x_{n+1})}{(1+ y_n x_{n+1})(1 + y_{n+1}x_{n+1})},\\ \nn \widetilde x_n &=& x_n\frac{(1 + y_n x_{n+1})(1 + y_{n-1}x_{n-1})}{(1+ x_{n+1})(1 + x_{n-1})}.
\end{eqnarray}
We consider each of these variables to be drawn from some statistical ensemble. Although the most general possibility has these being conditional probability assignments, a rich structure is nevertheless presented when restricting to cases where they are independently drawn from site-wise independent distributions.\\  

We denote this set of independent probability distribution functions as $\{\psi_n(y_n)\}$ for the $y_n$ variables, and $\{\chi_n(x_n)\}$ for the $x_n$ variables. Given that $\widetilde y_n = \widetilde y_n(y_n,y_{n+1},x_{n+1})$, and $\widetilde x_n = \widetilde x_n(x_{n-1},x_n,x_{n+1},y_{n-1},y_n)$, it must be the case that under decimation, successive RG transformations induce a flow on the probability assignments given by 
\begin{align}
\label{renY}\widetilde\psi_n(\widetilde y_n \equiv y) &= \int_{0}^\infty\hspace{-10pt}...\int_{0}^\infty \hspace{-5pt}dx_{n+1}dy_{n}dy_{n+1}\, \psi_n(y_n)\\\nn &\times \psi_{n+1}(y_{n+1})\chi_{n+1}(x_{n+1})\delta(\widetilde y_n - y),
\end{align}
and
\begin{align}
\hspace{-10pt}\label{renX}\nn&\widetilde \chi_n(\widetilde x_n \equiv x) = \int_{0}^\infty\hspace{-10pt}...\hspace{-3pt}\int_{0}^\infty\hspace{-5pt}dx_{n-1}dx_{n}dx_{n+1}dy_{n-1}dy_{n} \delta(\widetilde x_n - x)\\ &\times \hspace{-2pt}\psi_{n-1}(y_{n-1}) \psi_n(y_{n})\chi_{n-1}(x_{n-1}) \chi_{n}(x_n) \chi_{n+1}(x_{n+1}),
\end{align}
where the dependent variables $\widetilde y_n$ and $\widetilde x_n$ that appear in the arguments of the delta functions are given by Eqs. \ref{Ybn}. The renormalized distributions are normalized to unity by default, as seen by integrating both sides of the above with respect to $y$ and $x$, respectively.\\ 

We first consider Eq. \ref{renY}, where we note that the delta function integral can be performed by changing the integration variable to $\widetilde y_n$. To do this, we must first invert $\widetilde y_n$ given by Eq. \ref{Ybn} in terms of the variable we look to transform, which we choose to be $y_n$. The result of this inversion is given by 
\eq{ynsol}{y_n = \frac{(1 + x_{n+1}) - (1 + y_{n+1}x_{n+1})\widetilde y_n}{\left[x_{n+1}(1 + y_{n+1}x_{n+1})\widetilde y_n - (1+ x_{n+1})y_{n+1}x_{n+1}\right]},}
so that after accounting for the Jacobian factor and performing the subsequent $\widetilde y_n$ integral, one ends up with
\begin{align}
\label{renY2}\nn
&\widetilde \psi_n( y) = \int_{0}^\infty\hspace{-5pt}\int_{0}^\infty\hspace{-5pt} dy_{n+1} dx_{n+1}\Bigl\{\Theta(y_n)\psi_n(y_n[x_{n+1},y_{n+1},y])\\ \nn &\frac{\psi_{n+1}(y_{n+1})\chi_{n+1}(x_{n+1})(1 + x_{n+1})(1 + y_{n+1}x_{n+1})|y_{n+1} - 1|}{x_{n+1}\left[(1 + y_{n+1}x_{n+1})y - (1+ x_{n+1})y_{n+1}\right]^2}\Bigr\},\\
\end{align}
where the Heaviside step function enforces the domain of integration to be such that Eq. \ref{ynsol} is positive for $y = \widetilde y_n$. Proceeding analogously for Eq. \ref{renX}, one obtains  
\begin{align}\nonumber 
\label{renX2}\widetilde \chi_n(\widetilde x_n \equiv x) &= \int_{0}^\infty\hspace{-10pt}...\int_{0}^\infty dy_{n-1}dy_{n} dx_{n-1} dx_{n+1}\Bigl\{\chi_n(x_n[...])\\& \nonumber \psi_{n-1}(y_{n-1})\psi_n(y_{n})\chi_{n-1}(x_{n-1})\chi_{n+1}(x_{n+1})\\ \times &\frac{(1+ x_{n+1})(1 + x_{n-1})}{(1 + y_n x_{n+1})(1 + y_{n-1}x_{n-1})}\Bigr\},
\end{align}
where the implicit dependence of $x_n$ on the remaining independent variables is given by:
\eq{xnsol}{x_n = \widetilde x_n\frac{(1+ x_{n+1})(1 + x_{n-1})}{(1 + y_n x_{n+1})(1 + y_{n-1}x_{n-1})}.}
One can either specify the probability assignments $\chi(x)$ and $\psi(y)$, or one can specify them for the original variables $C(\varepsilon)$ and $P(g)$ via the relations:
\begin{align}\psi(y) &= P(g[y])\frac{dg}{dy} \equiv \frac{P(g[y])}{y},\\ \nn \chi(x) &= C(\varepsilon[x])\frac{d\varepsilon}{dx} \equiv \frac{C(\varepsilon[x])}{x},
\end{align}
where we have used that $g = \log y$ and $\varepsilon = \log x$. Repeating the same steps that led to Eqs. \ref{renY2} and \ref{renX2}, we obtain the induced RG flow for $P(g)$ and $C(\varepsilon)$ as
\begin{eqnarray}
	\nn
	\widetilde P_n(&&\widetilde g_n \equiv g) = \int_{-\infty}^\infty\int_{-\infty}^\infty\hspace{-5pt} dg_{n+1} d\varepsilon_{n+1}\Bigg\{\Theta(g_n) P_{n+1}(g_{n+1})\\ && \nn \frac{e^g(1 + e^{\varepsilon_{n+1}})(1 + e^{g_{n+1} + \varepsilon_{n+1}})|e^{g_{n+1}} - 1|}{e^{\varepsilon_{n+1}}\left[(1 + e^{g_{n+1} + \varepsilon_{n+1}})e^g - (1+ e^{\varepsilon_{n+1}})e^{g_{n+1}}\right]^2}\\ \label{renY22} && \times C_{n+1}(\varepsilon_{n+1})P_n(g_n[\varepsilon_{n+1},g_{n+1},\widetilde g_n \equiv g])\Bigg\},
\end{eqnarray}
where $g_n$ in the argument of $P_n$ above is obtained by taking the log of Eq. \ref{ynsol} and substituting the definitions for the $y$ and $x$ variables. Similarly, one finds:  
\begin{eqnarray}\nn
	&&\widetilde C_n(\widetilde \varepsilon_n \equiv \varepsilon) = \int_{-\infty}^\infty\hspace{-10pt}...\int_{-\infty}^\infty\hspace{-10pt} dg_{n-1}dg_{n} d\varepsilon_{n-1} d\varepsilon_{n+1}\Bigg\{P_{n}(g_{n})\\ \nn 
	&&P_{n-1}(g_{n-1})C_{n-1}(\varepsilon_{n-1})\frac{(1+ e^{\varepsilon_{n+1}})(1 + e^{\varepsilon_{n-1}})}{(1 + e^{g_n + \varepsilon_{n+1}})(1 + e^{g_{n-1} + \varepsilon_{n-1}})}\\ \label{renX22}
	&&\hspace{-5pt}\times C_{n+1}(\varepsilon_{n+1})C_n(\varepsilon_n[g_{n-1},g_n,\varepsilon_{n-1},\varepsilon_{n+1},\widetilde \varepsilon_n \equiv \varepsilon])\Biggr\}.
\end{eqnarray}
Although the convolutions in both sets of Eqs. \ref{renY2}, \ref{renX2} and Eqs. \ref{renY22}, \ref{renX22} are amenable to numerical integration for arbitrary probability assignments, it is informative to consider certain special cases.

\subsection{Sparse disorder}
\begin{figure}[t]
	\centering
		\includegraphics[width=0.95\linewidth]{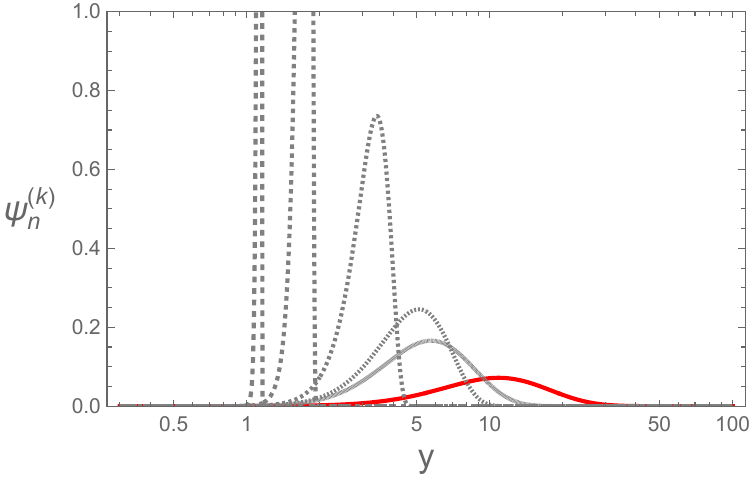}
		\label{fig:sub1}
		\includegraphics[width=0.95\linewidth]{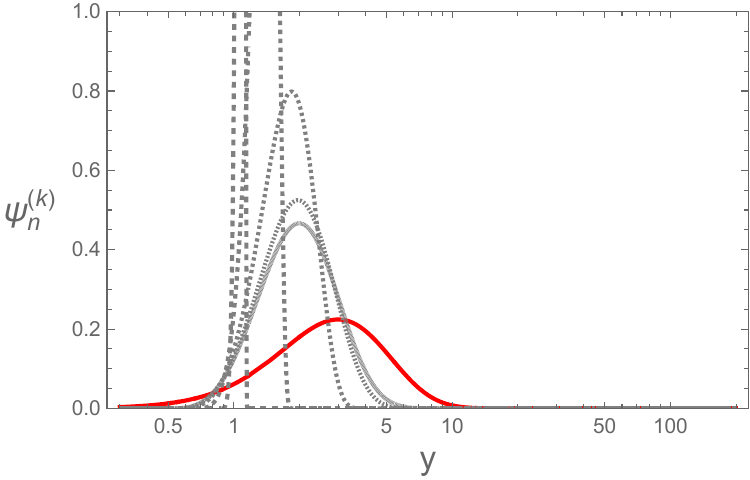}
		\label{fig:sub2}
	\caption{Top panel: Renormalization group induced deformation of the bare probability assignment $\Pi(y)$ given by Eq. \ref{bareP} for $\alpha_\pi = 5$, $\lambda_\pi = e^{-1}$ (red line) for $k$ from 1 to 5 (gray dashed lines with successively coarser dashing). Bottom panel: The same, but for $\Pi(y)$ given with parameters $\alpha_\pi = 4$, $\lambda_\pi = e^{0}$.}
	\label{fig:3}
\end{figure}

The simplest non-trivial example one could consider is the case where disorder is localized to a single site $n$, so that the bare probability assignments for the $x$ and $y$ variables are given by:
\begin{align}
	\label{SBD}
	\{\chi^{(0)}_a(x_a)\}_{a \neq n}&= \delta(x_a - x^{(0)}),~~ \chi^{(0)}_n(x_n) = \Omega(x_n)\\ \nonumber
	\{\psi^{(0)}_a(y_a)\}_{a \neq n}&= \delta(y_a - y^{(0)}),~~ \psi^{(0)}_n(y_n) = \Pi(y_n)
\end{align}
where $\Omega(x_n)$ and $\Pi(y_n)$ are the bare distributions of the $x$ and $y$ variables for the site where the disorder is localized.  Inserting these into Eqs. \ref{renY2} and \ref{renX2} yield the following induced renormalization group transformations on the probability assignments upon iteration:
\begin{align}\label{RGPDF1}
\chi^{(k+1)}_n(x) &=\int_0^\infty dy\Bigg\{ \frac{\psi^{(k)}_n(y)\left(1 + x^{(k)}\right)^2}{\left(1+ y x^{(k)}\right)\left(1+ y^{(k)} x^{(k)}\right)}\\ \nonumber &\times \chi^{(k)}_n\left[\frac{x\left(1+x^{(k)}\right)^2}{\left(1+ y x^{(k)}\right)\left(1+ y^{(k)} x^{(k)}\right)}\right]\Bigg\},\\ \label{RGPDF2}
\psi^{(k+1)}_n(y) &=\frac{(1 + x^{(k)})|y^{(k)}-1|(1+ x^{(k)}y^{(k)})}{x^{(k)}(y + y^{(k)}[x^{(k)}(y-1) - 1])^2}\\ \nonumber &\times\psi_n^{(k)}\left[\frac{1 + x^{(k)} - y(1 + x^{(k)}y^{(k)})}{x^{(k)}(y[1 + x^{(k)}y^{(k)}] - y^{(k)}[1+x^{(k)}])}\right]_+\hspace{-3pt},
\end{align}
where dummy variables have been relabeled for compactness, and where
\begin{align}
\label{RGPDF3}
\{\chi^{(k+1)}_a(x)\}_{a \neq n} &= \delta(x - x^{(k+1)}),\\ \{\psi^{(k+1)}_a(y)\}_{a \neq n}&= \delta(y - y^{(k+1)}), \nn
\end{align}
where the arguments of the delta functions $x^{(0)}$ and $y^{(0)}$ in Eq. \ref{SBD} get recursively renormalized to $x^{(k)}$ and $y^{(k)}$ via Eqs. \ref{Ybn} as:
\begin{eqnarray}
	\label{YbnH}
	y^{(k+1)} &=& \frac{\left(1 + y^{(k)^2}x^{(k)}\right)\left(1 + x^{(k)}\right)}{\left(1+ y^{(k)} x^{(k)}\right)^2},\\ \nonumber x^{(k+1)} &=& x^{(k)}\frac{\left(1 + y^{(k)} x^{(k)}\right)^2}{\left(1+ x^{(k)}\right)^2}.
\end{eqnarray}
The effects of successive RG transformations are particularly straightforward to evaluate for the link disorder assignment $\psi^{(k)}_n$ given that no convolutions are involved in Eq. \ref{RGPDF2}. The subscript on Eq. \ref{RGPDF2} is to denote an implicit factor of the Heaviside step function that enforces the positivity of the argument of $\psi_n^{(k)}$.\\

For illustrative purposes, we choose the bare distributions for both $x$ and $y$ to be Gamma distributed over the interval $[0,\infty)$, so that 
\begin{eqnarray}
	\label{bareP}
\Pi(y) &=& \frac{\lambda_\pi^{\alpha_\pi}}{\Gamma(\alpha_\pi)} y^{\alpha_\pi - 1} e^{-\lambda_\pi y },\\ \nn
\Omega(x) &=& \frac{\lambda_\omega^{\alpha_\omega}}{\Gamma(\alpha_\omega)} x^{\alpha_\omega - 1} e^{-\lambda_\omega x },
\end{eqnarray}
where the mean and the variance, given by $\alpha/\lambda$ and $\alpha/\lambda^2$ respectively. which one can specify arbitrarily. Note that the $x$ variables relate to a probability assignment for a corresponding Erd\"os-Renyi random graph via the relation (cf. Eq. \ref{hdef}):
\eq{}{x = e^\varepsilon = \frac{p}{1-p}.}
Upon repeated decimation transformations, we see that the disorder probability assignment would asymptote towards a delta function distribution centered around $y = 1$ in Fig. \ref{fig:3}, corresponding to the $g = 0$ fixed line in Fig. \ref{fig:RG}. This is not particularly surprising, as the latter is a global attractor, and every realization within the disorder ensemble would flow towards this fixed line under repeated decimation.\\

One can also adapt these results to the case where one has disorder localized to multiple, but sufficiently sparsely situated sites\footnote{In more precise terms, this would be localized disorder that vanishes in the thermodynamic limit.}. The renormalization of the local probability assignments would proceed according to Eqs. \ref{RGPDF1} and \ref{RGPDF2} up until the RG step where any two sites where disorder has been assigned would coalesce under decimation. At this step, one has to revert to the expressions for the general transformation Eqs. \ref{renY2}, \ref{renX2}, iterating again from there with Eqs. \ref{RGPDF1}, \ref{RGPDF2} until the next step where coalescence occurs (cf. \cite{Bar-Yam:2018ytq}).

\subsection{Uniform disorder}

A more general scenario would be to consider the $\{y_a\}$ and $\{x_a\}$ variables to each be independently and identically distributed. Although the general convolutions expressed in Eqs. \ref{renY2} and \ref{renX2} can be iterated in this case, it is useful to recast the former in simplified and more symmetrical form by transforming to the variables $z_n, x_n$, where $z := x_n y_n = e^{\varepsilon_n + g_n}$, so that Eqs. \ref{Ybn} transform to: 
\begin{eqnarray}
	\label{Zbn}
	\widetilde z_n &=& \frac{(x_n + z_n z_{n+1})(1 + z_{n-1})}{(1+ x_{n-1})(1 + z_{n+1})},\\ \nonumber \widetilde x_n &=& \frac{(x_n + z_n x_{n+1})(1 + z_{n-1})}{(1+ x_{n+1})(1 + x_{n-1})}.
\end{eqnarray}
We now presume the disorder to be uniformly assigned as: 
\begin{align}
	\label{UBD}
	\{\chi^{(0)}_n(x_n)\}_{\forall n }& = \chi(x_n),~~~~~~	\{\psi^{(0)}_n(z_n)\}_{\forall n }= \psi(z_n),
\end{align}
for bare probability distributions $\chi(x), \psi(z),$ where the latter is given by the product distribution for the variables $x$ and $y$, which can in principle be specified independently. Repeating the same steps that led to Eqs. \ref{renY2} and \ref{renX2}, one arrives at the following induced RG transformations for the probability assignments at any given site\footnote{Here, and in all analogous derivations that follow, one writes down the analogous equations to Eqs. \ref{renY} and \ref{renX}, identifies a convenient integration variable which admits a unique inverse for the RG transformations for this variable alone analogous to Eqs. \ref{Ybn}, and performs the delta function integration accounting for the appropriate Jacobian factor and integration limits.}:
\begin{eqnarray}
	\label{renX3}&&\hspace{-10pt}\chi^{(k+1)}(x) = \int_{0}^\infty\hspace{-12pt}...\int_{0}^\infty\hspace{-5pt} du\, dv\, dy\, dw\Bigg\{\frac{(1+ w)(1 + v)}{(y + uw )}\\ \nn &&\hspace{-10pt}\psi^{(k)}\hspace{-3pt}\left[\frac{x(1 + w)(1 + v)}{(y + w u)}-1 \right]_+\hspace{-5pt}\psi^{(k)}(u)\chi^{(k)}(v)\chi^{(k)}(y)\chi^{(k)}(w)\Bigg\},
\end{eqnarray}
and
\begin{eqnarray}\nn
	&&\psi^{(k+1)}(z) = \int_{0}^\infty\hspace{-12pt}...\int_{0}^\infty\hspace{-5pt} du\, dv\, dy\, dw\Bigg\{ \psi^{(k)}(v)\psi^{(k)}(y)\psi^{(k)}(w)\\ && \label{renX4} \hspace{-10pt} \frac{(u+ yw)(1 + v)}{z^2(1 + w)}\chi^{(k)}\hspace{-3pt}\left[\frac{(u+ yw)(1 + v)}{z(1 + w)}-1 \right]_+\hspace{-5pt}\chi^{(k)}(u)\Bigg\},
\end{eqnarray}
where again, the $+$ subscripts denote an implicit Heaviside theta function in the integrand to ensure the positivity of the arguments of $\psi^{(k)}$ and $\chi^{(k)}$ in the above. The two equations represent a set of coupled non-linear convolutions that can be iterated repeatedly to evaluate the induced renormalization group flow on the bare probability assignments $\chi^{(0)}$ and $\psi^{(k)}$. Their evaluation can be simplified in some cases by noticing that both convolutions can be expressed as  
\begin{eqnarray}
	\label{renX5}\chi^{(k+1)}(x) = \frac{d}{dx}&& \int_{0}^\infty\hspace{-10pt}...\int_{0}^\infty\hspace{-5pt} du\, dv\, dy\, dw\Bigg\{\Psi^{(k)}\left[f(x|u,v,y,w)\right] \nonumber \\ && \times \psi^{(k)}(u)\chi^{(k)}(v)\chi^{(k)}(y)\chi^{(k)}(w)\Bigg\},
\end{eqnarray}
\begin{eqnarray}
\nonumber
	\label{renX6}\psi^{(k+1)}(z) = -\frac{d}{dz}&& \int_{0}^\infty\hspace{-10pt}...\int_{0}^\infty\hspace{-5pt} du\, dv\, dy\, dw\Bigg\{\Xi^{(k)}\left[g(z|u,v,y,w) \right] \nn \\ && \hspace{-7pt} \times\chi^{(k)}(u)\psi^{(k)}(v)\psi^{(k)}(y)\psi^{(k)}(w)\Bigg\}.
\end{eqnarray}
where $\Psi^{(k)}(s)$ is the cumulative distribution function $\Psi^{(k)}(s) = \int_0^s dt\, \Theta(t)\psi^{(k)}(t)$ where the step function under the integrand enforces the vanishing of $\Psi$ where the integrand has no support, where $\Xi^{(k)}(s)$ is defined analogously, and where the respective functional arguments are defined as: 
\begin{eqnarray}
f(x|u,v,y,w) &:=& \frac{x(1 + w)(1 + v)}{(y + w u)}-1,\\ \nn
g(z|u,v,y,w) &:=& \frac{(u+ yw)(1 + v)}{z(1 + w)}-1.
\end{eqnarray}

Eq. \ref{renX5} has the advantage of recasting the induced renormalization group flow on the probability assignments as a gradient flow. 
\begin{figure}[t]
	\centering
		\includegraphics[width=0.95\linewidth]{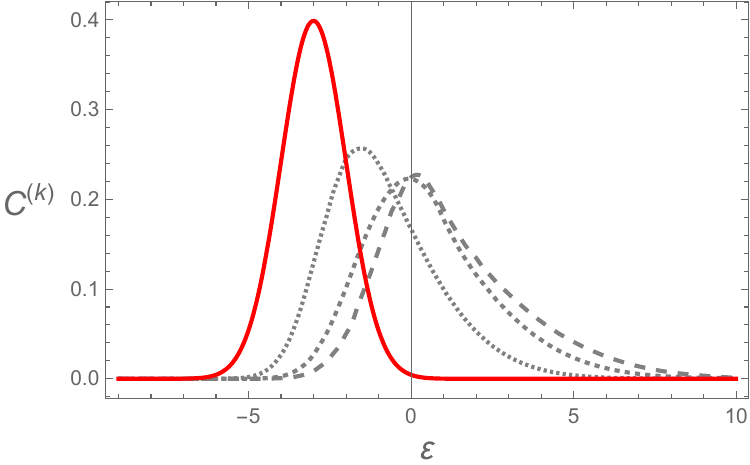}
		\label{fig:sub3}
		\includegraphics[width=0.95\linewidth]{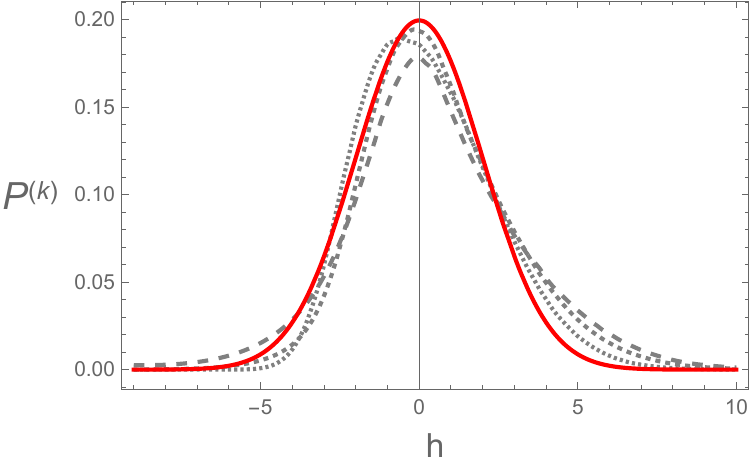}
		\label{fig:sub4}
	\caption{Renormalization group induced deformation of the uniform probability assignments $C^{(k)}(\varepsilon)$ (top) and $P^{(k)}(h)$ (bottom). Bold red lines denote the bare probability assignments, taken to be Gaussians with means and variances given by $\mu_\varepsilon = -3, \sigma_\varepsilon = 1$ and $\mu_h = -3, \sigma_h = 1$, respectively, and the gray dashed lines represent successive RG convolutions running from $k=1$ to $3$ with increasingly coarse dashing.}
	\label{fig:uniform}
\end{figure}
One can also repeat the derivation to obtain the analogous expressions for the distribution of $\varepsilon$ and the joint distribution for $h = \varepsilon + g$, denoted $C^{(k)}(\varepsilon)$ and $P^{(k)}(h)$ respectively:
\begin{eqnarray}\nonumber
	\hspace{-15pt}\label{renX7}C^{(k+1)}(\varepsilon) =&& \int_{-\infty}^\infty\hspace{-10pt}...\hspace{-3pt}\int_{-\infty}^\infty\hspace{-12pt} du\, dv\, dy\, dw\Bigg\{\left[1 - \frac{(e^y + e^{u+w} )}{e^\varepsilon(1+ e^w)(1 + e^v)}\right]^{-1}_+\\ \nn &&\hspace{-10pt}\times P^{(k)}\left[\log\left(\frac{e^\varepsilon(1 + e^w)(1 + e^v)}{(e^y + e^{w+ u})}-1\right) \right]\\ &&\hspace{-10pt} \times P^{(k)}(u)C^{(k)}(v)C^{(k)}(y)C^{(k)}(w)\Bigg\},
\end{eqnarray}
and
\begin{eqnarray}\nonumber
	\hspace{-18pt}\label{renX8}P^{(k+1)}(h) =&& \int_{-\infty}^\infty\hspace{-10pt}...\hspace{-3pt}\int_{-\infty}^\infty\hspace{-10pt} du\, dv\, dy\, dw\Bigg\{\left[ 1 - \frac{(1 + e^w)e^h}{(e^u+ e^{y+w})(1 + e^v)}\right]_+^{-1}\\ \nn &&\hspace{-20pt} \times C^{(k)}\left[\log\left(\frac{(e^u+ e^{y+w})(1 + e^v)}{(1 + e^w)e^h}-1 \right)\right]\\&&\hspace{-20pt} \times C^{(k)}(u)P^{(k)}(v)P^{(k)}(y)P^{(k)}(w)\Bigg\},
\end{eqnarray}
where the subscripts now on the Jacobian factors again restrict the integration domain to be such that the respective factors are both positive. As before, one can recast both convolutions as an analogous gradient flow:
\begin{eqnarray}\nn
	C^{(k+1)}(\varepsilon) = \frac{d}{d\varepsilon}&& \int_{-\infty}^\infty\hspace{-10pt}...\int_{-\infty}^\infty\hspace{-10pt} du\, dv\, dy\, dw\, \Bigg\{\Pi^{(k)}\left[F(\varepsilon|u,v,y,w) \right]\\ \label{renX9} && \times P^{(k)}(u)C^{(k)}(v)C^{(k)}(y)C^{(k)}(w)\Bigg\},
\end{eqnarray}
\begin{eqnarray}\nn
	P^{(k+1)}(h) = -\frac{d}{d h}&& \int_{-\infty}^\infty\hspace{-10pt}...\int_{-\infty}^\infty\hspace{-10pt} du\, dv\, dy\, dw\Bigg\{\Xi^{(k)}\left[G(h|u,v,y,w)\right]\\ && \times C^{(k)}(u)P^{(k)}(v)P^{(k)}(y)P^{(k)}(w)\Bigg\},\label{renX10}
\end{eqnarray}
where $\Pi^{(k)}(s) = \int_0^s dt\, \Theta(t)P^{(k)}(t)$, and $\Xi^{(k)}(s) = \int_0^s dt\, \Theta(t)C^{(k)}(t)$, where we've defined:
\begin{eqnarray}
	F(\varepsilon|u,v,y,w) &:=& \log\left\{\frac{e^\varepsilon(1 + e^w)(1 + e^v)}{(e^y + e^{w+ u})}-1\right\}\hspace{-2pt},\\ \nn
	G(h|u,v,y,w) &:=& \log\left\{\frac{(e^u+ e^{y+w})(1 + e^v)}{(1 + e^w)e^h}-1 \right\}\hspace{-1pt}.
\end{eqnarray}
We illustrate the induced RG flow on the bare uniform disorder probability assignments in Fig. \ref{fig:uniform}. Visual inspection of both Figs. \ref{fig:3} and \ref{fig:uniform} seems to suggest the induced renormalization group flow to be akin to some sort of active transport process for the disorder probability assignments. This can in fact be made precise as we do in appendix \ref{sec:CD}, where we identify this as time reversed drift-diffusion, which is simply the cumulative result of pointwise deformations of the PDFs generated by the beta functions that generate the flow.

\section{Implications and applications}

Over the preceding chapters, we have shown how restricting, or truncating the effective Hamiltonian to quadratic order (cf. Eq. \ref{eq:ham}) results in closed form renormalization group transformations whose repeated iterations result in pairwise interactions flowing towards increasing irrelevance. For infinite, or very large networks, one would say that such interactions flow to \textit{triviality} at large scales. We defer the discussion as to whether this feature persists once higher order interactions are introduced to our concluding remarks, and discuss the implications of these facts for various applications in what follows. We do this by first distilling our findings to their bare elements, and then contextualizing them in a number of applied settings. These elements are:
\begin{itemize}
	\item Disordered Erd\H{o}s-Renyi random graphs with pairwise conditioning of the individual links admit closed form renormalization group transformations for maximum coordination number two. 
	\item Repeated RG transformations render these interactions irrelevant, flowing towards a disordered non-interacting Erd\H{o}s-Renyi random graph.  
	\item Were one to draw the couplings of the effective Hamiltonian from a disorder ensemble, the effects of pairwise conditioning results in localized skewing of the macroscopic probability assignments relative to their microscopic (bare) assignments.  
	\item The induced RG flow on the disorder probability assignment corresponds to time reversed drift-diffusion on the associated statistical manifold. 
\end{itemize}
We now detail the ramifications of these elements for a number of applications. What follows is only a partial survey, with further applications, especially when higher order conditionings are incorporated and suggestive of future lines of inquiry revisited in our concluding remarks. 

\subsection{Peer, preferential, susceptibility, and reinforcement effects on random networks}
Random graphs are often invoked as a mathematical model for social networks, contact networks, or relationship networks more generally. Allowing the probabilities to form links depending on graph properties associated to a given node can lead to an variety of interesting network structures and topologies. Preferential attachment models \cite{Baraba_si_1999, Albert_2002} are a widely studied class of network growth models that associate the probability of nodes sequentially added to a graph to connect with existing nodes in a manner that depends on their degree (see e.g. \cite{newman2018networks, piva2020} for a survey). This is implemented via an attachment kernel, where the link formation probability is given by $p_{i} = f(k_i)/\sum_j f(k_j)$, where the sum is over the existing nodes, and where $f(k_i)$ is some function of the degree $k_i$ of the nodes. Within the context of a generalized statistical mechanical formalism, preferential attachment models are possible to realize for linear attachment kernels provided we go beyond maximum coordination number two. This can be seen by the fact that any conditioning of a given element of the adjacency matrix on the degree of a node that it might connect to would necessitate operators of the form: 
\eq{eq:lin}{	H(\textbf{A}) \supset A_{ik}\sum_{j\neq k} A_{ij}.} 
Such terms would correspond to terms in an effective Hamiltonian of the general form Eq. \ref{LineHam} as opposed to Eq. \ref{GH} on the corresponding line graph representation. Non-linear attachment kernels of the form $p_i \propto f(k_i)$ would necessitate the addition of a series of higher order interactions of the form
\eq{eq:nlin}{	H(\textbf{A}) \supset A_{ik}\Bigl(\sum_{j\neq k} A_{ij}\Bigr)^n}
that depend on the Taylor expansion of the kernel function $f(k_i)$, for which the bilinear truncation Eq. \ref{LineHam} would no longer be fit to purpose. Although for unweighted networks, the fact that the $A_{ij} = \{0,1\}$ implies that expressions of the form Eq. \ref{eq:nlin} can always be brought into the form Eq. \ref{eq:lin}, the resulting expressions will not be readily translatable back into a function of the degree, which could hinder inference problems that utilize the corresponding operator expansion in the effective Hamiltonian. That, along with the fact that operators of the form Eq. \ref{eq:nlin} are unavoidable for any generalization to unweighted networks necessitates their consideration in generality.\\

It is, of course, incumbent upon any complete treatment to make precise the manner in which existing classes of random graph models transcribe into the generalized statistical mechanical formalism, which we expand upon in our concluding remarks. In what follows, we wish to proceed unencumbered by precedent and explore how qualitatively similar peer and preferential effects can naturally be modeled within the class of bilinear graph effective Hamiltonians. In order to do this, it first helps to visualize the action of the different operators that make up the effective Hamiltonian on a given graph.\\
\begin{figure}[t]
	\includegraphics[width=0.8\linewidth]{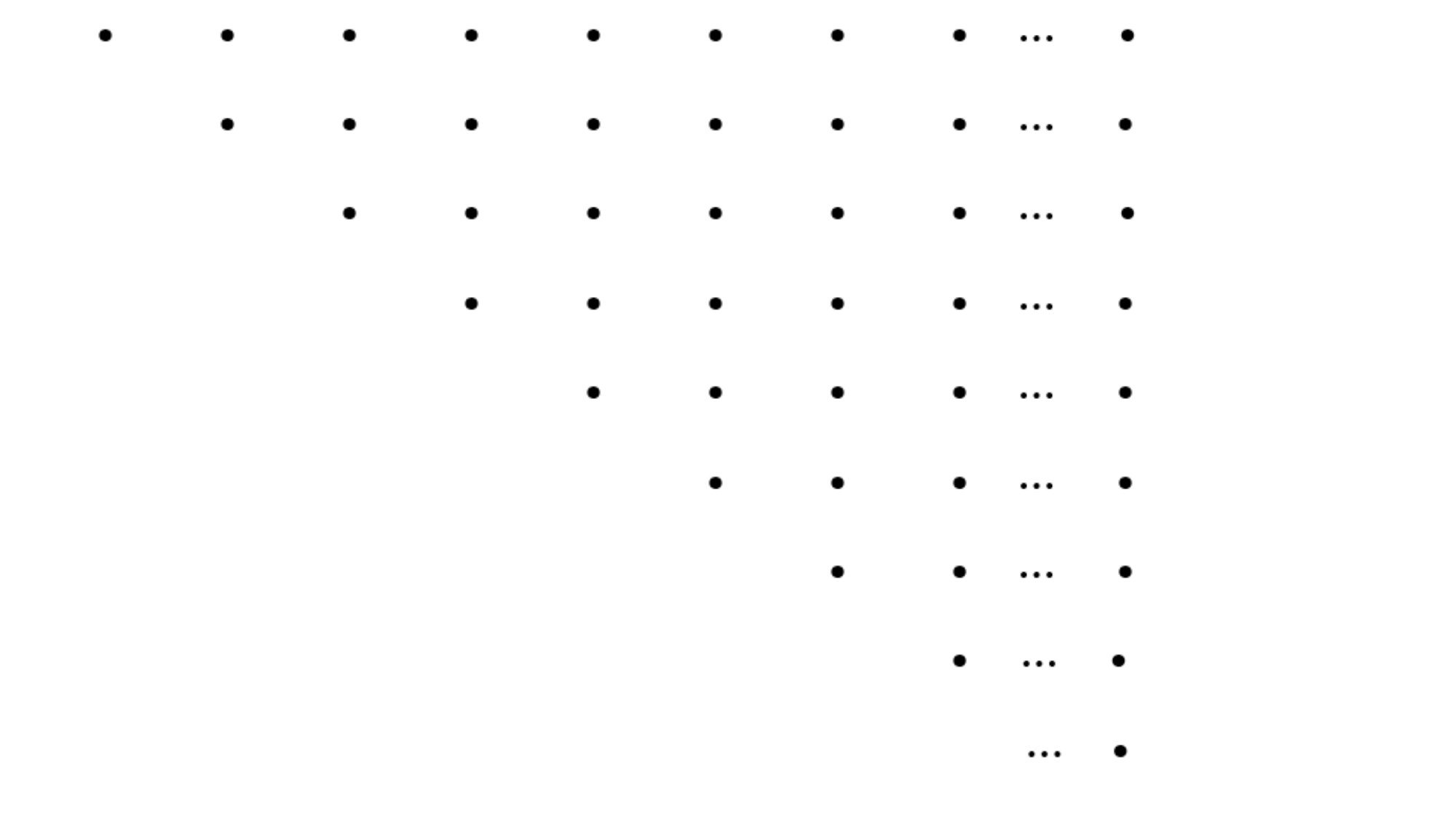}\\
	\vspace{10pt}
	\includegraphics[width=0.8\linewidth]{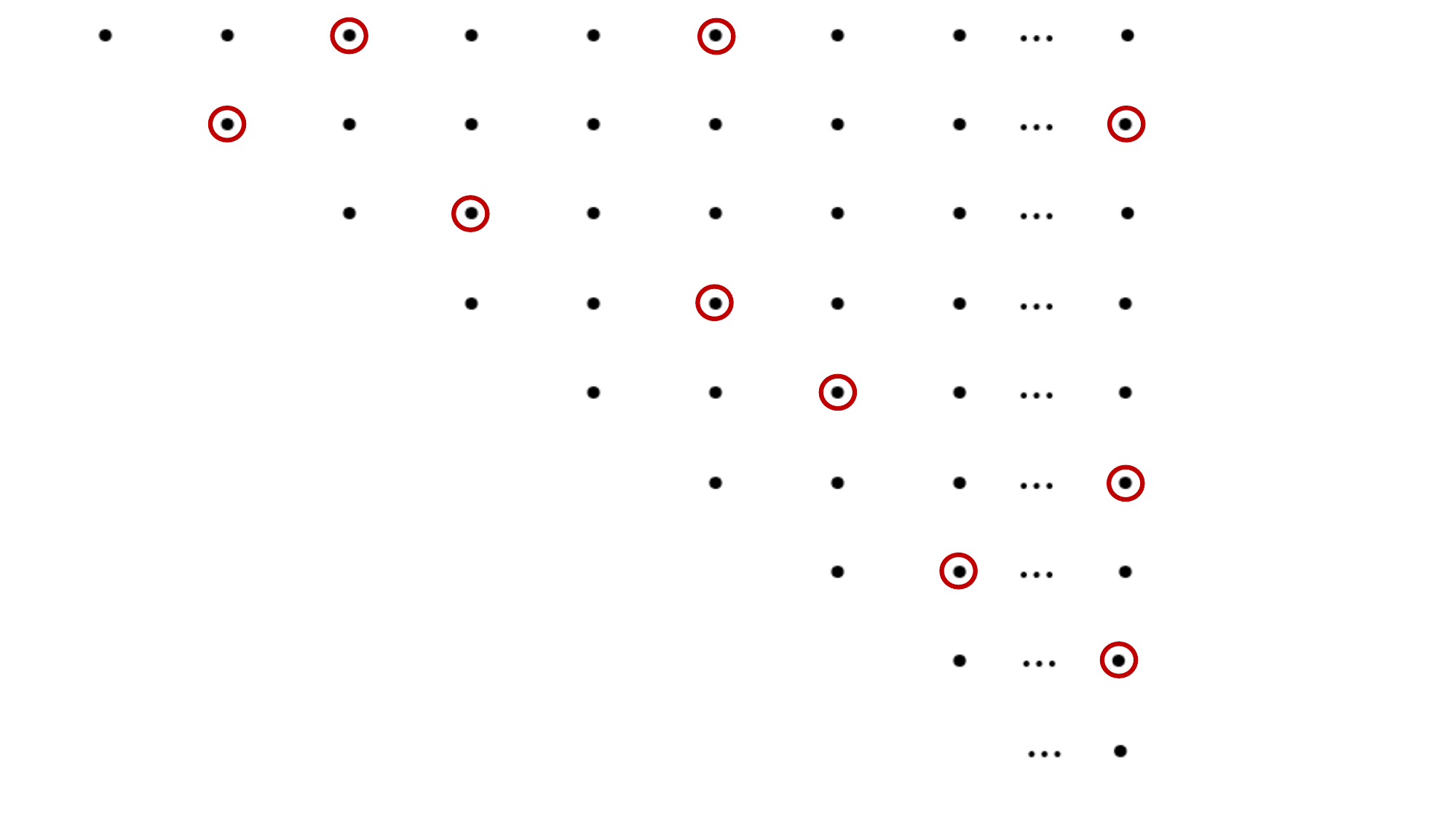}
	\caption{Schematic representation of a random graph, where each dot represents a possible link (i.e. an off diagonal entry of the adjacency matrix). The realization of a given link in a specific draw corresponding to the relevant entry of the adjacency matrix are denoted by circles.}
	\label{fig:cn1}
\end{figure}

\begin{figure}[t]
	\centering
	\includegraphics[width=0.8\linewidth]{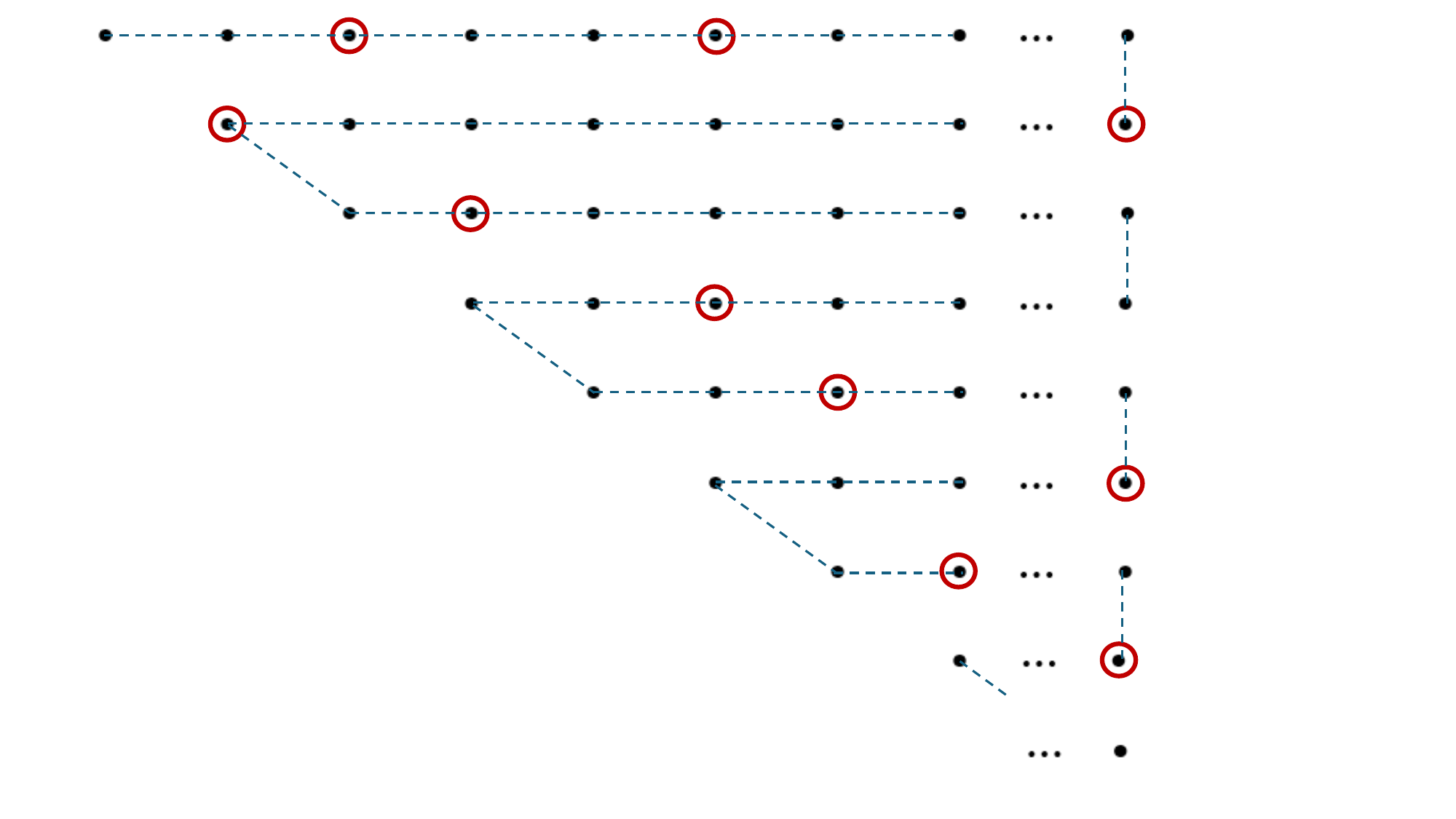}\\
	\vspace{10pt}
	\includegraphics[width=0.8\linewidth]{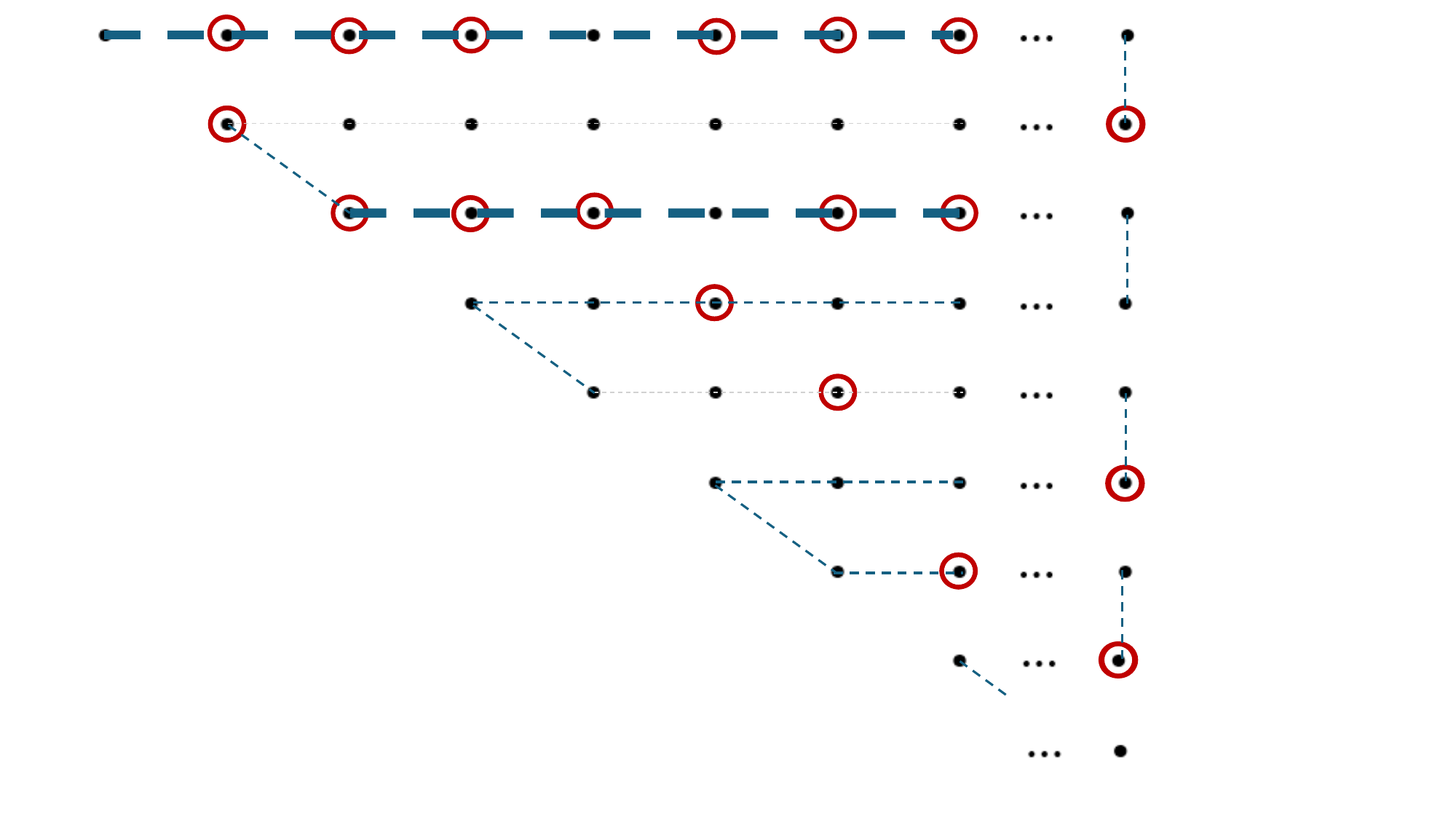}
	\caption{Representation of a random graph with maximum coordination two interactions (top panel). Enhanced or diminished attachment and preference effects can be implemented via the sign and magnitude of the relevant bilinear couplings $\{g_a\}$ (bottom panel: bold and faint lines, respectively).}
	\label{fig:cn2}
\end{figure}

In Fig. \ref{fig:cn1} we schematically represent a random graph with a fixed maximum number of nodes via the elements of the associated adjacency matrix, with each site representing a possible link\footnote{\label{fn:o}With the understanding that the ordering is inconsequential and the following discussion applicable to an arbitrary ordering.}. In this representation, a link realized in a given draw is denoted by a circle, whose probabilities are determined by the set of $\{\varepsilon_{a}\}$ in Eq. \ref{GH}, with the remaining unconnected nodes discarded. We represent the maximum coordination number two pairwise interactions between links (i.e. the set of $\{g_a\}$) as dashed lines that couple the sites in Fig. \ref{fig:cn2}, caveating footnote \ref{fn:o}.  From this figure, we see that we can actualize preferential attachment in a complimentary manner by assigning relatively larger positive couplings for all pairwise interactions that index a specific node, depicted with bold lines in the bottom panel of Fig. \ref{fig:cn2}. We can also realize preferential  \textit{avoidance} via negative values for the bilinear couplings that index a particular set of nodes. The random assignment and distribution of nodes that have either preferential or avoidant conditioning can be implemented via a suitable choice for the disorder ensemble from which the couplings are drawn.\\    

One can see immediately infer other ways in which one can induce qualitatively similar preferential effects on an ensemble of random graphs without invoking pairwise conditioning. One could, for instance, set all the $\{g_a\}$ to be vanishing and specify the resulting disordered Erd\H{o}s-Renyi probability assignments, i.e. the set of $\{\varepsilon_a\}$, to be arranged so that the desired distribution of degrees is attained, as implemented in \cite{PhysRevE.70.066117, Garuccio_2023, Servedio_2004, van2024random}. In more precise terms, one can in principle reproduce a fixed number of summary statistics that derive from correlation functions of the graph degrees of freedom up to a given order from a continuous family of effective Hamiltonians. This degeneracy is a feature common to all effective theories, whereby one needs to determine up to a certain number of quantities in order to specify the couplings of the effective Hamiltonian completely. The degeneracy gets broken once one demands that a sufficient number of additional summary statistics be reproduced by the effective Hamiltonian, typically receiving contributions from higher order correlation functions, such as local clustering coefficients and the number and distribution of triangles, for example.\\

One might also wonder how decimation on the line graph (where we decimate over links rather than nodes) maps back onto a random graph representation. Given that the marginalization effected on the line graph by each decimation reduces the number of link degrees of freedom by a factor of two, the result of two consecutive decimations would be to render a graph with $\mathcal N = N(N-1)/8$ links, which in the limit of large graphs corresponds to the number of links for a graph with $\widetilde N = N/2$ nodes\footnote{There will be a number of hermit nodes that will result from this process, whose number go as $1/N$.}. One can therefore view the ensemble of networks generated by the effective Hamiltonian obtained by decimation on the line graph after an even number of RG transformations to admit a random network representation with $\widetilde N = N/2^m$ nodes, for some power $m$. As we have stressed in the introduction, this is by no means the only possibility as one can generate arbitrary RG transformations through the composition of transformations that decimate over individual links or sites. A more comprehensive treatment of this would fall within the ambit of a broader study of scheme dependence when renormalizing effective graph Hamiltonians, which deserves a separate detailed investigation.\\

The scaling behavior of preferential effects can be inferred from Figs. \ref{fig:RG}, \ref{fig:3}, and \ref{fig:uniform}, all of which illustrate the progressive weakening and eventual vanishing of pairwise link conditioning at large enough scales. This is of course, when restricted to maximum coordination number two interactions, which merely transcribes the infinite temperature fixed point on a one dimensional disordered Ising spin chain onto a random graph model, as discussed at the end of section \ref{sec:HRG}. One can translate the implications of this into various applications straightforwardly.\\

Although utilized through some variant in a number of applications that model the structure and diffusion processes over social, contact, or population networks, it is widely appreciated that Erd\H{o}s-Renyi random graphs by themselves fail to accurately model a variety of real-world networks. This is in part due to the Poisson nature of their degree distributions, which is seldom observed in application, necessitating additional structural requirements or restrictions \cite{6856208, PhysRevE.89.012116, alizadeh2017generating, SHI200733, 6937219, PhysRevE.85.056109}. One variant of particular utility is a random graph model of Molloy and Reed, which is constructed to reproduce an arbitrarily specified degree distribution \cite{Molloy}, which neveretheless still comes up short in modeling the distance distributions observed in number of data sets of a given degree distribution \cite{doi:10.1073/pnas.012582999}. Whether this is because of biased estimation in the given data sets, or because of unknown relations among elements of the graph which the model fails to capture is something that the effective generalized statistical mechanical framework offers the means to address and rectify, by allowing for either the possibility of arbitrary pairwise conditioning, or the possibility of modeling data limitations via assigning distribution priors on the relevant couplings.\\

In the context of social or contact networks, Fig. \ref{fig:RG} offers a useful mnemonic for the scaling behavior of individual couplings encoded in Eq. \ref{RGg} even if it restricts to the case of homogeneously assigned couplings. One sees that any localized positive or negative conditioning results in skewing the probabilities for random links to appear at macroscopic scales towards greater or lesser values, respectively. Positive conditioning can model, for instance, peer pressure or preferential effects on social networks, where the existence of a link between any two nodes makes it likelier for other links to form, or the converse. Similarly for contact networks, the probability that any two agents have been in contact with one another are certainly not independent, and positively conditioned to varying and inhomogeneously assigned degrees in general. Our findings show that such conditioning will eventually result in a disordered Erd\H{o}s-Renyi assignment at the most coarse-grained level. One could therefore conceive of utilizing this conditioning in the context of temporal networks to model susceptibility effects in the context of opinion dynamics and reinforcement effects in the context of artificial and biological neural networks.\\ 

Dynamical processes on networks come in two broad categories. One can either take the underlying random graph as a fixed background on which dynamical processes occur, or one can view the structure of the graph itself to evolve over time. Both processes can also occur simultaneously and codependently. By assigning state variables to every node, one can associate a variety of dynamical structures with these variables that will depend on the network structure. Diffusion processes, for example, are naturally modeled via the graph Laplacian. More generally, one can model arbitrary dynamical processes via operators constructed from the adjacency and degree matrices, as the graph analog of the derivative expansion Eq. \ref{eq:diff}. One might also consider the structure of the network to evolve over time as a temporal Markov process on a layered network, where each layer represents the state of the network at any given time step. Dynamical processes that interpolate between the layers can be derived from the properties of the graph structure, or state variables at the nodes, where coordination number two conditioning between links in adjacent layers can model can model arbitrary temporal correlations between layers.\\ 

In this way, the generalized statistical mechanical formalism developed here can readily be adapted towards classifying and studying the behavior of temporal and spatial correlations and conditioning, and their effects on late time or terminal summary statistics (such as the outcomes of elections) in opinion dynamic models based on random graphs \cite{albi2016, 7577815, GHADERI20143209, PhysRevE.74.036112, 10092193, Schoenebeck, Emenheiser, dedomenico2025peerinfluencebreaksergodicity}. Similarly, reinforcement effects at the cortical level in biological and artificial neural networks \cite{doi:10.34133, PMID:19404458, doi:10.34133, ALKAM2025100220, PhysRevE.105.044312, doi:10.1073pnas.2316745122,  Mocanu_2018, arya2025}, when represented as layered temporal networks can also be modeled via pairwise, disordered, and inhomogeneously assigned pairwise conditioning, although any meaningful applications of relevance would almost certainly have to go beyond the maximum coordination number two approximation invoked in this investigation.

\subsection{Network inference with limited data}

\begin{figure}
	\centering
	\includegraphics[width=\linewidth]{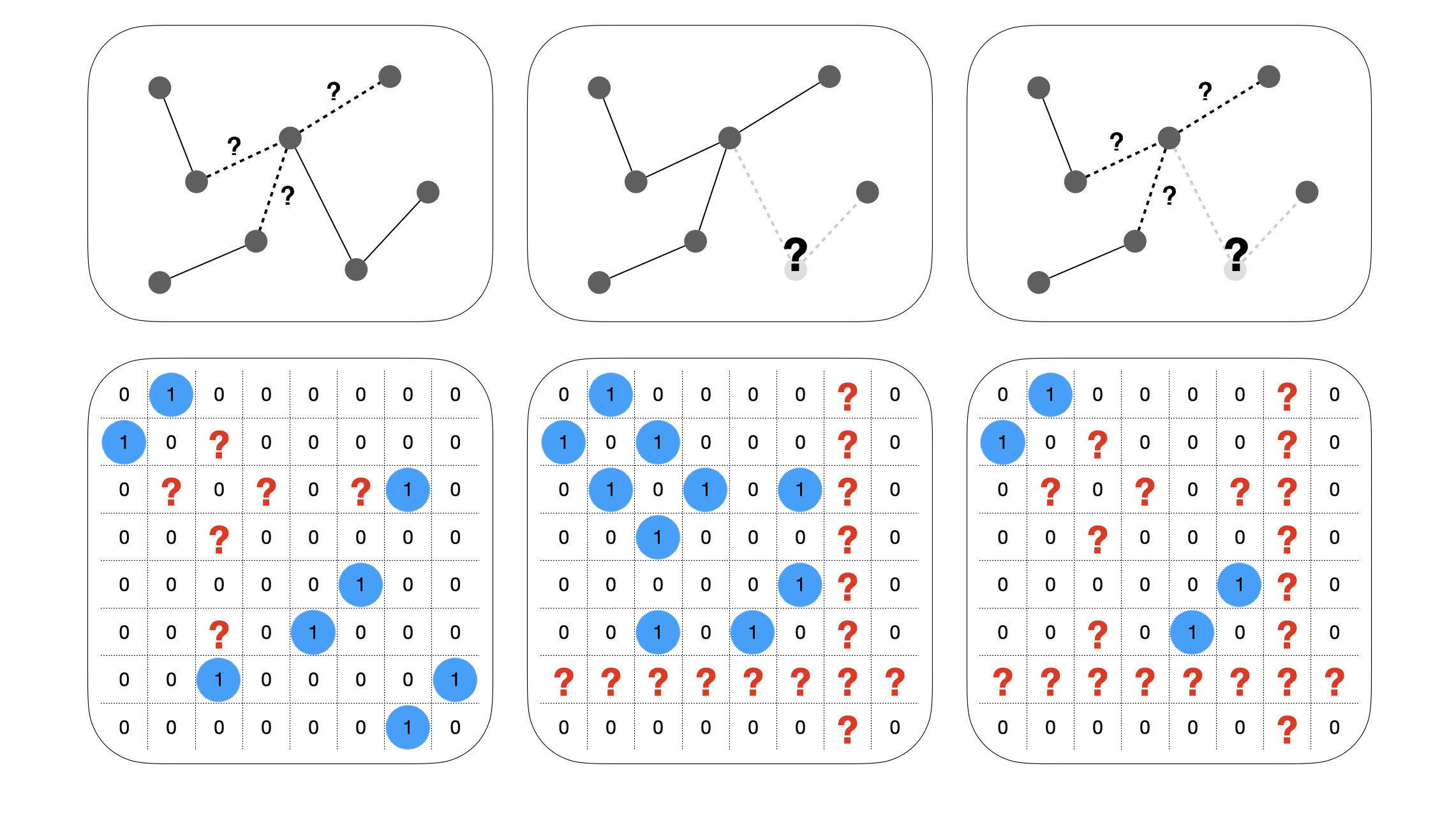}
	\caption{Schematic depiction of incomplete network data. From left to right: individual links are missing or unresolvable; individual nodes are either unobserved, unresolvable, or of unknown connectivity; both possibilities simultaneously.}
	\label{fig:possibleuncertainty}
\end{figure}

Yet another application of the framework developed in the preceding sections constitutes something of an inversion of the usual manner in which RG methods are utilized, which is towards inference and reconstruction problems \cite{cimini2021reconstructing, squartini2018reconstruction} in the context of limited data resolution. The workhorse for this derives from the observation that in any line graph representation of a random network, decimation renormalization group transformations are formally equivalent to the process of statistical marginalization over the degrees of freedom that have been integrated out (cf. appendix \ref{app:A}, caveating \ref{app:NM}), which can also be viewed as marginalizing over variables we are unable to access or measure. This seemingly simple paraphrasing opens a portal for addressing how one can infer or reconstruct network representations with limited data. The most basic instance of this is when we simply have no access to, or have unreliable or noisy estimators for a finite subset of graph variables in attempting to determine whether a link exists between any two nodes, or whether any two links are conditioned on each other. In statistical inference, one proceeds by viewing these as `nuisance parameters', where by assigning prior probability assignments to model the uncertainty, and subsequently marginalizing over these parameters, one arrives at an effective description for the remaining observable degrees of freedom. This same process would translate into a local renormalization transformation on a graph representation that incorporates these unknown or inaccessible graph parameters via corresponding uncertainty priors.\\

A more sophisticated application of the machinery developed in the previous sections would go further, and consider the partial local inversion utilized in preceding sections to try and `solve' for the family of underlying, or `fine-gained' parameters in terms of those that are inferred or measured at macroscopic scales. As we discuss in a companion paper \cite{Nextpaper}, in situations where one has closed RG equations -- as we do in the limited context of maximum coordination number two quadratic effective graph Hamiltonians -- and where all parameters of a network are known except for those corresponding to a given link and its corresponding conditioning, one obtains two equations in two unknowns that can be locally inverted, allowing us to `reconstruct' the missing information up to the accuracy of graph data at the observed scale and the extent to which higher coordination number conditioning can be neglected.\\ 

If one is dealing with a system where the RG equations do not close, as will be the case when higher order coordination numbers and interactions are incorporated (which will be the case in the majority of practical applications), additional higher-order interactions will be generated under RG flow. On the other hand, the transformations themselves restrict the set of parameters that are consistent with renormalization group flow to have mapped to the accessible parameters. In this case, sub-manifolds in parameter space can be defined as level curves of the RG transformation equation with the observed, or macroscopic parameters. One can then impose additional conditions and priors to further restrict the `fine-grained' sub-manifold consistent with what is observed. Given that higher-order couplings often run to irrelevance under certain conditions, one has heuristic restrictions on parameter space dictated by the resolution at hand. The error one is willing to accept on the higher-order terms not being identically zero, coupled with the respective RG equation for that parameters become equations that determine the subspace of microscopic parameters consistent with what is observed, or inferred. A more in depth expansion on this approach can be found in \cite{Nextpaper}, where we furthermore apply this formalism to economic and trade data of limited resolution.

\section{Concluding remarks and future directions}

The work presented in the preceding sections represents a first attempt towards addressing the question of what aspects of interactions, conditioning, and correlations across a complex network are relevant for its behavior at macroscopic scales. A more complete treatment -- necessary if any of the insights gained are to apply across the breadth of possible applications -- would have to incorporate higher order conditioning and coordination numbers in its analysis. To do this in full generality is of course intractable, but the subtle yet highly advantageous shift in operational perspective offered by the paradigm of effective theories focuses this problem into the more tractable question of classifying parameter (ir)relevance in any effective generalized statistical mechanical formalism. By allowing for higher order and coordination number interactions in the effective Hamiltonian Eq. \ref{eq:ham}, one can conceive of proceeding via perturbative techniques to explore the flow of couplings within certain truncations.\\ 

For a limited class of interactions that represent parametrized deviations from the maximum coordination number two case studied in the previous sections, one can be assured of the irrelevance of such higher order and higher coordination number couplings from analogous results on lattice systems. On the other hand, it will certainly be the case that novel and qualitatively different phenomena will arise when one no longer assumes higher order couplings to be perturbatively small. Nevertheless, any incremental gains in understanding the parameter relevance and possible phase structure of higher order and coordination number interactions will be of great use in any attempt to classify universality classes of effective graph Hamiltonians. It is to be stressed that within the generalized statistical mechanical framework, extensions to weighted and directed graphs can readily be accommodated via the operator expansion Eq. \ref{eq:ham} and the configuration space integrated or summed over.\\

Another important application, if not consistency check on the generalized effective statistical mechanical formalism would be to construct and identify the explicit operator expansions that correspond to existing network models. However, perhaps the most important direction to pursue from the present vantage would be towards applications of topical relevance. Even with the preliminary results for pairwise coordination number two interactions, a number of extensions towards models reliant on random graph representations was partly surveyed in previous sections, and will be elaborated upon in further detail in a followup investigation \cite{Nextpaper}.

\begin{acknowledgments}
This publication is part of the project ``Network renormalization: from theoretical physics to the resilience of societies’’, with file number NWA.1418.24.029 of the research programme NWA L3 - Innovative projects within routes 2024, which is (partly) financed by the Dutch Research Council (NWO) under the grant https://doi.org/10.61686/AOIJP05368, and the project ``Redefining renormalization for complex networks’’, with file number OCENW.M.24.039 of the research programme Open Competition Domain Science Package 24-1, which is (partly) financed by the Dutch Research Council (NWO) under the grant https://doi.org/10.61686/PBSEC42210.

This work is also supported by the European Union - NextGenerationEU - National Recovery and Resilience Plan (Piano Nazionale di Ripresa e Resilienza, PNRR), projects ``SoBigData.it - Strengthening the Italian RI for Social Mining and Big Data Analytics'', Grant IR0000013 (n. 3264, 28/12/2021) (\url{https://pnrr.sobigdata.it/}), and ``Reconstruction, Resilience and Recovery of Socio-Economic Networks'' RECON-NET EP\_FAIR\_005 - PE0000013 ``FAIR'' - PNRR M4C2 Investment 1.3.

Work by AC is also supported by \#NEXTGENERATIONEU (NGEU) and funded by the Italian Ministry of University and Research (MUR), National Recovery and Resilience Plan (NRRP), project MNESYS (PE0000006) – A Multiscale integrated approach to the study of the nervous system in health and disease (DN. 1553 11.10.2022). 

SP wishes to thank Paolo Benincasa, La Ricotta Proyect and the Asociación Cultural La Ricotta for hospitality over the writing of this manuscript as well as many useful discussions, for which he is also grateful to Kevin Grosvenor.\\
\end{acknowledgments}	

\appendix

\section{\label{app:A} Generating functionals, and when renormalization $\equiv$ marginalization}

Given any set of $\mathcal N$ random variables denoted $X = \{x_1,...x_{\mathcal N}\}$ that are drawn from an ensemble defined by the probability measure $P(x_1,...,x_{\mathcal N})$, all statistical moments can be obtained via the corresponding generating functional $Z[J]$, defined as
\eq{}{Z[J] = \int d^{\mathcal N}x\, P(x_1,...,x_{\mathcal N})\, e^{-(J,x)},}
where the scalar product $(J,x) := \sum_{i=1}^{\mathcal N} J_i x_i$. From this definition, it follows straightforwardly that 
\eq{}{\langle x_1^{a_1}... x_{\mathcal N}^{a_{\mathcal N}}\rangle =  (-1)^M\frac{\partial^M Z[J]}{\partial J_{1}^{a_1}\partial J_{2}^{a_2}... \partial J_{\mathcal N}^{a_{\mathcal N}} }\Bigg|_{J\equiv 0}}
where we've defined $M := \sum_{i=1}^{\mathcal N} a_i$, and furthermore require $\int d^{\mathcal N}x\, P(x_1,...,x_{\mathcal N}) = 1$. Similarly, given a probability measure $P(\textbf{A})$ defining an ensemble of $\mathcal N\times \mathcal N$ real matrices, where    
\eq{pm}{\int \mathcal D \textbf{A} P[\textbf{A}] = 1; ~~~~\mathcal D\textbf{A} := \prod_{i,j = 1}^{\mathcal N} d A_{ij},}
one can define the analogous generating functional \cite{Jackson:2021kpp}:
\eq{}{Z[J] = \int \mathcal D\textbf{A}P[\textbf{A}]\, e^{-{\rm Tr\,}A J^T}.}
Expectation values of arbitrary powers of the matrix \textbf{A} can be obtained from successive derivatives of $Z[J]$
\eq{moments}{\langle \textbf{A}^m_{ij} \rangle = (-1)^m\frac{\delta^m Z[J]}{\delta J_{ii'}\delta J_{i'k}... \delta J_{l l'}\delta J_{l'j} }\Bigg|_{J\equiv 0}.}
Were one to consider the random matrix ensemble to correspond to adjacency matrices of a random graph, additional restrictions on the probability measure will be required depending on whether the network is weighted or not, or directed or not. Unweighted, undirected graphs correspond to symmetric matrices with entries that are either vanishing, or unity. For example, the Gilbert-Erd\H{o}s-Renyi model \cite{Erdos2022OnRG, gilbert1959random} for $N$ nodes where links can appear independently with probability $p$ would have an associated probability measure given by
\begin{eqnarray}
	\label{G}
	(i < j)~~~~~~ P_{ij} &=& p\,\delta(A_{ij} - 1) + (1-p)\,\delta(A_{ij})\\ \nn
	(i = j)~~~~~~ P_{ij} &=& \delta(A_{ii})\\ \nn 
	(i > j)~~~~~~ P_{ij} &=& \delta(A_{ij} - A_{ji}).
\end{eqnarray}
Since the probability measure factorizes into identical independently drawn probabilities, the generating functional $Z[J]$ is straightforwardly evaluated as
\begin{eqnarray}\label{zans}Z[J] &=&  \prod_{i,j}\int\,d A_{ij}\, P_{ij}e^{- A_{ij}J_{ij}}\\ \nn &=& (1-p)^{\frac{N(N-1)}{2}}\prod_{i < j}\left(1 + e^{\left[\varepsilon - J_{(ij)}\right]} \right),
\end{eqnarray}
where we've defined
\eq{GZ}{\varepsilon := \log\left(\frac{p}{1-p}\right),~J_{(ij)} := J_{ij} + J_{ji}\ .}
One can also define the generating function for all connected correlation functions $W[J]$ as
\eq{}{e^{-W[J]} := Z[J],} 
which is given by
\eq{GW}{W[J] = \frac{N(N-1)}{2}\log\left(\frac{1}{1-p}\right) - \sum_{i < j} \log\left(1 + e^{\left[\varepsilon - J_{(ij)}\right]} \right).}
Discarding the source independent term above (which corresponds to discarding the source independent normalization factor in Eq. \ref{zans}), one obtains 
\begin{eqnarray}
\label{GZ0}Z[J] &=& \prod_{i < j}\left(1 + e^{\left[\varepsilon - J_{(ij)}\right]} \right),\\ \nn W[J] &=& - \sum_{i < j} \log\left(1 + e^{\left[\varepsilon - J_{(ij)}\right]} \right).
\end{eqnarray}
Identifying $W[J]|_{J\equiv 0}$ as the free energy associated with the corresponding exponential random graph model, one sees that it is equivalent to the thermodynamic free energy of a gas of non-interacting fermions at zero chemical potential localized at the links 
\eq{FE}{W[J]|_{J\equiv 0} = -\sum_{i < j} \log\left[1 + e^{\varepsilon_{ij} } \right],}
where the above is the generalization to the case where the individual link probabilities $p$ vary across the graph (i.e. $p \to p_{ij}$ in Eq. \ref{G}). This is particularly manifested when one evaluates the link occupation number:
\eq{}{\langle \textbf{A}_{ij} \rangle = -\frac{1}{Z}\frac{\delta Z}{\delta J_{ij}}\Bigg|_{J\equiv 0} \hspace{-10pt} \equiv n_{ij} = \frac{1}{e^{\varepsilon_{ij}} + 1} = p_{ij},}
where one makes the identification 
\eq{}{\varepsilon_{ij} \equiv \frac{(E-\mu)_{ij}}{k_BT}.} 
By taking additional derivatives of $Z[J]$ with respect to the $J_{ij}$, one can calculate expectation values of any observable derivable from the adjacency matrix. For example, the average number of paths of length $k$ connecting nodes $i$ and $j$, defined as $\#_{ij}(k)$ is given by:
\eq{}{\#_{ij}(k) = \langle \textbf{A}^k_{ij} \rangle,}
and the expectation value of the degree of the $i^{th}$ node $\langle k_i\rangle$, or the average degree of the entire network $\langle k \rangle$ can be obtained as
\eq{eq:deg}{\langle k_i \rangle =  \sum_{j=1}^N \langle \textbf{A}_{ij} \rangle, ~\langle k \rangle =  \frac{1}{N}\sum_{i,j=1}^N \langle \textbf{A}_{ij} \rangle,}
and so on. The analogs of the above expressions for general heterogeneous networks can also be found, for instance in \cite{cimini2019statistical, Squartini_2015}.

Consider now the situation where we have a partition of our set of random variables into two disjoint subsets $X = \{\textbf{y}, \textbf{x}\}$, where without loss of generality, we order them such that $\{\textbf{y}\} := \{x_1,...,x_k\}$ and $\{\textbf{x}\} := \{x_{k+1},...,x_{\mathcal N}\}$, with joint probability distribution given by $\Pi(\textbf{y},\textbf{x})$. We furthermore presume that we either have no means of accessing or measuring the \textbf{x} degrees of freedom, or we are simply uninterested in computing any statistical moments that involve them. In that case, any observable $\mathcal O(\textbf{y})$ that is constructed exclusively out of the \textbf{y} variables can be re-expressed as
\begin{eqnarray*}
	\langle \mathcal O(\textbf{y})\rangle &=& \int d\textbf{y} d\textbf{x}\, O(\textbf{y})\Pi(\textbf{y},\textbf{x})\\ &=& \int d\textbf{y} \, O(\textbf{y})\int d\textbf{x}\, \Pi(\textbf{y},\textbf{x}) : =  \int d\textbf{y} \, O(\textbf{y})\Pi_{\rm eff}(\textbf{y}),
\end{eqnarray*}
where 
\eq{marg}{\Pi_{\rm eff}(\textbf{y}) := \int d\textbf{x}\, \Pi(\textbf{y},\textbf{x})}
In the context of statistical inference, this corresponds to the well worn procedure of marginalization. When viewed from the vantage of all correlation functions  being generated by a moment generating functional (or partition function), where the probability measure can be expressed as a (graph) Hamiltonian via the equivalence $\Pi(\textbf{y},\textbf{x}) = e^{-\beta H(\textbf{y},\textbf{x})}$, Eq. \ref{marg} becomes
\eq{eq:eff}{\int d\textbf{x}\, e^{-\beta H(\textbf{y},\textbf{x})} := e^{-\beta'H_{\rm eff}(\textbf y)},} 
which is formally the process of computing probabilities with an effective Hamiltonian that has integrated out the $\{\textbf{x}\}$ degrees of freedom.\\ 

In  the context of physical systems, the resulting effective probability measure Eq. \ref{eq:eff} goes by various names depending on the criteria used to separate the $\{\textbf{x}\}$ and $ \{\textbf{y}\}$ variables. In the case where $\{\textbf{x}\}$ represents short wavelength (or high energy) degrees of freedom that we have no access to, and the $ \{\textbf{y}\}$ variables represent the long wavelength degrees of freedom that we construct all observable quantities out of, the resulting coarse grained probability measure goes by the name of the (Euclidean) \textit{Kadanoff-Wilson effective action}. In certain cases, one might instead be interested in integrating out, or marginalizing over \textit{all} modes of a given field or set of degrees of freedom regardless of their energetics. In the context of statistical field theory, such an object would be known as the \textit{one particle irreducible effective action} \cite{Helias_2020}. In both cases, the formally exact implementation of renormalization can be viewed as the moment generating functional equivalent of statistical marginalization.

\section{\label{app:NM}When renormalization $\not\equiv$ marginalization}
\begin{figure}[t]
	\centering
	\includegraphics[width=0.9\linewidth]{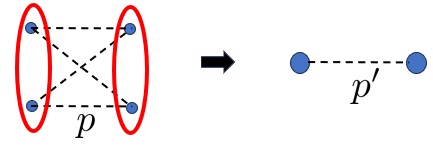}
	\label{fig:ER}
	\caption{Schematic depiction of a `block-averaging' renormalization of an Erd\H{o}s-Renyi random graph, where each successive RG step involves grouping nodes into blocked pairs, where the renormalized probability is defined as the probability that there exists at least one link connecting nodes across blocks.}
\end{figure}
The process of separating degrees of freedom by some criteria, and integrating out the subset designated to be marginalized over can be iterated repeatedly. Each step constitutes a renormalization group transformation, and corresponds to coarse graining if the separation criteria is a resolution scale\footnote{Which of course, presumes prior notions of metric distance, which may not always be available in the context of random networks.}. Although the complete execution of this procedure as described would indeed be equivalent to statistical marginalization, there are a number of situations where this will not be the case. For instance, it may be desirable to redefine the variables after every RG step, so that after each transformation, we have:
\eq{eq:cv}{\textbf{y}^{(n)} \to \textbf{y}'^{(n)}[\textbf{y}^{(n)}].}
Such transformations may be necessitated in order to render subsequent integrations of the form Eq. \ref{eq:eff} to be tractable. They may also implement redefinitions in to `coarse grained' variables (such as the block averaged variables familiar from spin systems) that one might deem to be more useful aggregate variables according to some criteria. In both cases, one is not forced to make such transformations when dealing with finite systems, which typically facilitate approximations whose artifacts one must take care to identify when determining macroscopic observables. The result of these transformations would be to induce a transformation on the effective Hamiltonian obtained at each step as
\eq{eq:ham2}{ H_{\rm eff}^{(n)}(\textbf{y}^{(n)}) \to \widetilde H_{\rm eff}^{(n)}(\textbf{y}'^{(n)})}
It is here that the non-trivial notion of \textit{scheme dependence} creeps into the discussion, which is perhaps best illustrated through example.\\

Consider a regular Erd\H{o}s-Renyi random graph defined by the probability measure Eq. \ref{G} with the associated generating functional Eq. \ref{zans}. Given the normalization of the probability measure for a finite graph with $N$ nodes, integrating out all links associated to a given node would simply return the probability measure for the links among the remaining $N-1$ nodes, which is also described by the measure Eq. \ref{G}, but omitting indices corresponding to the node that has been integrated out. In this case, the result of each successive RG transformation is the trivial mapping
\eq{eq:marg}{p \to p' = p ~~~~\mathrm{(marginalization)}}
On the other hand, one could instead coarse grain a specific draw of the ensemble into `block averaged' variables via the prescription depicted in Fig. \ref{fig:ER}. In this case, the probability of there being a link between any of the links within the blocked nodes is given by one minus the probability that no links appear. Therefore:
\eq{BA}{p \to p'=  1 - (1-p)^4 ~~~~\mathrm{(block~averaging)}}
which has an unstable fixed point at $p = 0$, and a stable fixed point at $p = 1$. It is clear that one gets a result different than in Eq. \ref{eq:marg} because of the non-trivial transformation induced by block averaging at each step.\\

On the other hand, one can also imagine averaging in blocks larger than two, as nothing forces us a priori to consider coarse graining degrees of freedom via a halving procedure. Repeating the steps that led to Eq. \ref{BA} for blocks of three nodes at a time, one obtains the transformation $p \to 1 - (1-p)^9$. More generally, and one obtains $p \to 1 - (1-p)^{m^2}$ when averaging over blocks consisting of $m$ nodes. In all cases regardless of the prescription, one finds the following fixed point structure: 
\begin{eqnarray*}
	p_* &=& 0 ~~~~(\mathrm{unstable})\\
	p_* &=& 1 ~~~~(\mathrm{stable})
\end{eqnarray*}
That is, although there is scheme dependence in the precise nature of the RG transformations and any intermediate quantities one may derive from them, \textit{the existence and values of the fixed points is identical in all schemes}. This is an example of a familiar phenomenon in particle physics effective theories, whereby scheme dependence itself is a diagnostic that one hasn't yet computed an unambiguously measurable quantity, rather, an artifact of one's approximations. Scheme \textit{independence} therefore provides a useful classification tool to identify quantities that represent legitimate macroscopic observables\footnote{For example, among the various quantities whose renormalization group running one may attempt to compute the running of in the context of continuum field theories, the form of the free energy (or effective potential) away from its critical points is generally scheme dependent, but the existence and values attained at critical points is not \cite{Andreassen:2014gha}. Furthermore, one can show that although the structure of divergent quantities that need regularization can be scheme dependent, the coefficients of logarithmic divergences, which dictate how physically observable quantities run with scale is scheme independent \cite{Negro:2024bbf}.}. In the context of the generalized statistical mechanics of random graph models, it is the existence of fixed points and the critical parameter values taken at them that represent such scheme independent quantities when block averaging.\\ 

Although we are under no prior compulsion to make successive transformations of the form Eq. \ref{eq:cv} in the context of finite systems, various approximation schemes, or the need to impose additional conditions derived from constraints on observational or data resolution may necessitate them in practice. On the other hand, in the context of very large systems, i.e. in the idealized limit where the number of nodes $N$ is taken to infinity (sometimes referred to as the thermodynamic limit), successive changes of variable, or `field redefinitions', are \textit{essential} in order to have a well defined framework to calculate probabilities. One of the reasons for this can immediately be inferred from the generalization of Eq. \ref{eq:eff} to a system with a continuum, or an infinite number of degrees of freedom:
\eq{eq:effC}{\int d\textbf{x}\, e^{-\beta H(\textbf{y},\textbf{x})} := \widetilde{\mathcal N} e^{-\beta'H_{\rm eff}(\textbf y)},} 
where $\widetilde N$ is the requisite normalization factor, that can itself be potentially divergent. Such a factor was already implicit in Eq. \ref{eq:eff} but equal to unity by construction, as the probability measure can always be taken for granted as appropriately normalized when dealing with a finite number of degrees of freedom.\\

One has to take additional care when confronted with infinite systems due to the potential divergences that arise can from the continuum or thermodynamic idealizations \cite{Burgess_2020, Helias_2020}. For present purposes, however, we simply note that the normalization in Eq. \ref{eq:effC} drops out of all observables derived from derivatives of the free energy $W = -\log Z$, and that successive \textit{re}normalizations can be absorbed by redefinitions of the remaining variables and the effective Hamiltonian at each step according to Eq. \ref{eq:cv} (wave function renormalization) and Eq. \ref{eq:ham2} (renormalization of couplings). These redefinitions themselves can be formally divergent. Nevertheless, a well worn prescription for calculating probabilities is still possible after a process that begins with first introducing an arbitrary \textit{regularization scale}. This can for instance, be a minimum resolution scale $\ell$, or a finite cutoff on the number of degrees of freedom $N$, where one takes $\ell \to 0$ or $N \to \infty$ at the end of the computation. The result will be formal divergences in either limit, which we then subtract with appropriate \textit{counterterms} in the effective Hamiltonian and measure factors. This process won't be without ambiguities, as subtracting infinities can result in differing finite remainders depending on the precise subtraction method. This is dealt with by fixing the finite remainder via demanding that the effective Hamiltonian reproduce the values of certain measurements at some fixed scale. This final step --- imposing \textit{renormalization conditions} --- only has to be done for a finite number of terms when working to a given order in parameter relevance.\\ 

The totality of the aforementioned steps --- regularization, subtraction, and the imposition of renormalization conditions constitutes the entire renormalization procedure in systems with an infinite number of degrees of freedom. After performing each one to completeness, one is left with a well defined formalism to unambiguously compute all relevant statistical moments, from which all observable quantities can be derived.

\section{\label{sec:CD} Induced RG flow as time reversed drift-diffusion}
Consider the space of couplings of an effective Hamiltonian that is closed under action of the renormalization group\footnote{The following can be generalized to situations where this assumption is relaxed with additional care.}. In the context of Eq. \ref{GH}, this would be the space $\mathbb R^{2\mathcal N}$, but more generally it could be an arbitrary manifold $\mathcal M$ (possibly with a boundary), with points that can be identified via local coordinates $\theta^a$, where $a$ runs from $1$ to $d$, the dimension of $\mathcal M$. The probability assignments for random draws of the adjacency matrix defined in Eq. \ref{eq:erg} implicitly depend on the couplings of the effective Hamiltonian, so that more explicitly:
\begin{equation}
	P(\textbf{A}|\theta) \equiv  \frac{e^{-H(\textbf{A}|\theta)}}{Z(\theta)},~~ Z(\theta) := \sum_{\{\textbf{A} \}} e^{-H(\textbf{A}|\theta)}. \label{eq:erg2}
\end{equation}
Each point in $\mathcal M$ parameterized by the coordinates $\theta^a$ corresponds to a specific effective Hamiltonian that generates random graphs according to the probability assignment above. $\mathcal M$ is therefore a \textit{statistical manifold}, where every point corresponds to a probability distribution function, which furthermore comes equipped with a natural metric tensor given by the Fisher-Rao metric \cite{rao1945information, e22101100, amari2000methods}:
\begin{eqnarray}\label{eq:im}
	G_{a b}(\theta) &=& \left\langle \frac{\partial \log P}{\partial \theta^a}\frac{\partial \log P}{\partial \theta^b} \right\rangle,\\  \nn &=& \sum_{\{\textbf{A} \}} P(\textbf{A})\frac{\partial \log P(\textbf{A}|\theta)}{\partial\theta^a}\frac{\partial \log P(\textbf{A}|\theta)}{\partial\theta^b}.
\end{eqnarray}
Although many interesting structures relating to parameter relevance and renormalization group flow can be inferred from and for the Fisher-Rao metric alone\footnote{In addition to the direct correspondence between the eigenvalues of the Fisher-Rao metric and RG parameter relevance discussed earlier \cite{Machta_2013}, one can also show that the metric undergoes a geometric flow  under coarse graining induced by the beta function vector field \cite{Raju_2018}. The information metric also geometrizes the study of the phase structure and critical exponents of a statistical mechanical system, as can be seen via Eqs. \ref{eq:erg2} and \ref{eq:im}, so that $G_{ab}(\theta) = - \partial_a\partial_b \log Z$. Recognizing the latter as the generalized free energy, we see immediately that singularities on the statistical manifold identify phase transitions \cite{Janke_2004, Dolan_2002, Erdmenger_2020}, and allows one to define and compute distances between theories \cite{balasubramanian2015}.}, it will play only a spectator role in what follows.\\

We are interested in computing the RG induced flow on the disorder probability assignments, which we represent by a probability measure over the statistical manifold itself. Denoting this measure as $\Pi(\theta) \equiv \Pi(\theta^1,\theta^2,...,\theta^n)$, the renormalization of the couplings $\theta^a$ induce an RG flow for $\Pi(\theta)$ according to  
\eq{eq:RGP}{D \Pi(\theta) = \beta^a\partial_a \Pi (\theta),}
where $\partial_a := \partial/\partial \theta^a$ and repeated indices summed over, and where we've defined the beta functions 
\eq{eq:tf}{D \theta^a = \beta^a(\theta).}
Here, the notation $D\theta^a$ is shorthand either for $D\theta^a := \partial \theta^a/\partial t$ where $t$ is the renormalization scale in the context of continuum systems\footnote{It is sometimes convenient to define continuum RG transformations as $\mu \,d\theta^a/d\mu$, where $\mu$ is some energy or length scale, in which case we identify $t/t_0 = \log \mu/\mu_0$, where the subscripted quantities refer to a reference scale, typically where renormalization conditions are imposed \cite{Burgess_2020, Negro:2024bbf} (cf. appendix \ref{app:NM}).}, or $D\theta^a := \theta^a_{k+1} - \theta^a_{k}$ for discrete RG transformations, where the subscript denotes the relevant RG step, with $\theta_0$ being the bare (i.e. initial) values of the couplings. In the present context, one can interchangeably consider the $\theta^a$ to be the $d = 2\mathcal N$ dimensional vectors $\theta^a := (\varepsilon_n,g_n), (x_n,y_n)$, or $(x_n,z_n)$, whose corresponding beta functions would follow from Eqs. \ref{RGg}, \ref{Ybn}, or \ref{Zbn} respectively. By recasting Eq. \ref{eq:RGP} as 
\eq{eq:WMD}{\left(D - \beta^a\partial_a\right) \Pi (\theta) = 0,}
one obtains the general solution
\eq{eq:gs}{\Pi(\theta) = F[\theta^a(\tau)],~~ F[\theta^a(0)] \equiv \Pi(\theta^a_0),}
where $F$ is any arbitrary function over $\mathcal M$, and where $\theta^a(\tau)$ is the solution to Eq. \ref{eq:tf} with initial conditions $\theta^a(0) = \theta^a_0$. In the above, $\tau$ denotes either the continuum variable $t$, or discrete variable $k$ that indexes the flow, and the bare probability assignment is denoted $\Pi(\theta_0)$. In the continuum case, the $\theta^a$ coordinatize the integral curves of the vector field generated by the beta functions.\\

The induced renormalization group flow on the probability assignments can be  straightforwardly interpreted via Eqs. \ref{eq:tf} and \ref{eq:gs} -- the bare PDF gets a point-wise functional deformation, with each point flowing along its integral curves defined by the respective beta functions. As seen from both Figs. \ref{fig:3} and \ref{fig:uniform}, the induced RG flow exhibits behavior akin to time reversed drift diffusion. This is in fact a general property of induced RG flow on the disorder probability assignments. In order to see this, we first make some preliminary observations, which for concreteness we frame in terms of the continuum limit\footnote{Which can also be viewed as the $N \to \infty$ limit in the discrete case, so that summations can be approximated by integrals and RG transformations can be iterated arbitrarily.}. The first is that conservation of probability under the action of RG transformations (cf. discussion below Eq. \ref{renX}) implies that 
\eq{eq:div}{\frac{\partial}{\partial t} \int_{\mathcal M}\hspace{-7pt}d^d\theta \, \Pi(\theta) = \int_{\mathcal M}\hspace{-7pt}d^{d-1}\theta\, \beta^a\partial_a\Pi(\theta)= 0,}
so that
\eq{eq:bc}{\da{\partial_a\beta^a} = \int_{\partial \mathcal M}\hspace{-10pt}d^{d-1}\theta\, \Pi(\theta)\, n_a \beta^a,}
where $n_a$ denotes a normal vector at the boundary of $\mathcal M$, and where double angled brackets denote an average over the disorder: 
\eq{eq:norm}{ \da{f(\theta)} := \int_\mathcal{M}\hspace{-7pt}d^{d}\theta\, \Pi(\theta) f(\theta).}
This is to be contrasted with averaging over the random variables, which we denote with single angled brackets as in Eq. \ref{eq:im}. For statistical manifolds without boundary, or whose boundary is at infinity (where the probability measure has to decay sufficiently rapidly to ensure a well defined normalization), the right hand side of Eq. \ref{eq:bc} vanishes so that 
\eq{eq:bc2}{\da{\partial_a\beta^a} = 0,}
which represents a globally (as opposed to locally) conserved disorder probability current. The lack of a locally conserved current is consistent with induced RG flows generally admitting local sources and sinks, corresponding to unstable and stable fixed points of the flow, even as disorder probability is globally conserved under action of the renormalization group.\\

At this stage, it behooves us to be more precise about the disorder probability measure. The freedom to reparameterize the disorder variables (as we have done freely in the previous sections), which is to allow for arbitrary coordinate transformations on $\mathcal M$, would require us to work covariantly and incorporate a volume factor $\sqrt{G}$ (with $G$ being the determinant of $G_{ab}$) into the measure, with all derivatives corresponding to covariant derivatives with metric compatible connections. Although the derivation that follows proceeds analogously for arbitrary connections, we restrict ourselves to flat connections over a specific coordinatization such that the measure is as indicated in Eqs. \ref{eq:div} and \ref{eq:norm}. This is not just to simplify the discussion, but also because a more comprehensive treatment would also admit more general possibilities for connections on the statistical manifold (e.g. the so-called one parameter family of $\alpha-$connections \cite{e22101100, amari2000methods}), which would deserve a separate discussion\footnote{In the present context, the probability assignments Eq. \ref{eq:erg2} correspond to an exponential family of distributions in the information geometric framework, and our choice to use the flat metric connections corresponds to working with an $\alpha-$connection with $\alpha = 1$ \cite{amari2000methods, e22101100}}.\\

Acting twice on the probability measure with the operator $\partial_t = \beta^a\partial_a$ results in the expression:
\eq{}{\partial_t^2 \Pi(\theta,t) = \left(\beta^c\partial_c\beta^a\right)\partial_a\Pi(\theta,t) + \beta^a\beta^c\partial_a\partial_c\Pi(\theta,t).} 
The term in the round parentheses can be recognized as the directional derivative of the beta function vector field tangent to itself, and can without loss of generality be decomposed as
\eq{}{\beta^c\partial_c\beta^a = \frac{\beta^a}{\kappa_\parallel} + \frac{\hat{v}_\perp^a}{\kappa_\perp}}

where $\hat v_\perp^a$ is defined to be a unit norm orthogonal drift vector that itself depends on $\theta$, as do $\kappa_\parallel$ and $\kappa_\perp$, where the latter can be viewed as analogous to the inverse radius of curvature of the vector flow of $\beta^a$ in a Frenet-Serret formalism \cite{frenet1852abstract, Serret1851}. As defined, $\kappa_\parallel$ will always be a positive non-zero number. With the decomposition above and recalling the definition $\partial_t = \beta^a\partial_a$, when one presumes the hierarchy $\partial_t\Pi \gg \kappa_\parallel\partial_t^2\Pi$ (or enforces it by rescaling the parameter $t$), one arrives at
\eq{eq:diff0}{-\frac{d }{dt}\Pi(\theta,t) = v_d^a\partial_a\Pi(\theta,t) + \kappa_\parallel \beta^a\beta^b\partial_a\partial_b\Pi(\theta,t)}
with $v_d^a := \frac{\kappa_\parallel}{\kappa_\perp}\hat v^a_\perp$, which formally describes time reversed anistropic drift-diffusion for the probability assignments under renormalization group flow.\\

We stress in closing the present discussion that in the context of continuum systems, the so-called Exact Renormalization Group implementation of Wilsonian renormalization \cite{wilson1974renormalization, Polchinski:1983, Bagnuls_2001,Berges_2002} can be also be recast as functional diffusion \cite{Cotler_2023, Berman_2023, Berman_2024} for probability assignments of the form $P(\textbf{x}) = e^{-H(\textbf{x})}/Z$ in analogy with Eq. \ref{eq:erg2}. In particular, Polchinsky's formulation of the ERG \cite{Polchinski:1983}, which can be viewed as a particular case of the Wegner-Morris flow equation \cite{Wegner1974, Morris_2000}, has a finite dimensional analog that can be recast as
\eq{eq:fd}{\frac{d}{dt}P (\textbf{x},t) = \partial_i \left[v^i P(\textbf{x},t)\right]+ g^{ij}(\textbf{x},t)\partial_i\partial_j P(\textbf{x},t)}
\\
where $v^i$ is drift term that depends of the particular regularization scheme \cite{Cotler_2023, Berman_2023} (cf. Appendix \ref{app:NM} for a recap of the renormalization procedure for continuum systems). Although Eq. \ref{eq:fd} apppears to be analogous to Eq. \ref{eq:diff0}, their physical content is very different as $P$ refers to the probability assignment for the stochastic degrees of freedom Eq. \ref{eq:erg2}, whereas $\Pi$ refers to the disorder probability assignment over the statistical manifold.

\end{document}